\documentclass[pra,twocolumn,superscriptaddress,nofootinbib,noshowpacs,preprintnumbers,longbibliography,floatfix]{revtex4-2}

\usepackage[utf8]{inputenc}
\usepackage{graphicx}
\usepackage{float}
\usepackage{amssymb}
\usepackage{amsmath}  
\usepackage{adjustbox}
\usepackage{mathtools}
\usepackage{dsfont}
\usepackage[overload]{empheq}
\usepackage{array}
\usepackage{bbold}
\usepackage{bm,fixmath}
\usepackage{mathrsfs}
\usepackage{pifont}
\usepackage{multirow}
\usepackage{upgreek}
\usepackage{xcolor}
\usepackage{bm}
\usepackage{bbm}
\usepackage{physics}
\usepackage{slashed}
\usepackage{tikz}

\usetikzlibrary{quantikz}

\usepackage[pdftex,
            pdftitle={Step scaling},
            pdfauthor={abc},
            bookmarks,
            colorlinks,
            linkcolor=myblue,
            citecolor=mymagenta,
            menucolor=black,
            urlcolor=myblue,
            plainpages=false,
            pdfpagelabels,
            hypertexnames=false]{hyperref}

\graphicspath{{Figures/}}
\definecolor{mymagenta}{RGB}{200, 0, 100}
\definecolor{myblue}{RGB}{45, 48, 146}
\definecolor{mypurple}{RGB}{200, 112, 255}

\usepackage{orcidlink} 

\begin{document}
\title{Towards determining the (2+1)-dimensional Quantum Electrodynamics \\ running coupling with Monte Carlo and quantum computing methods}

\author{Arianna~Crippa \orcidlink{0000-0003-2376-5682}}
\email{arianna.crippa@desy.de}
\affiliation{
 CQTA, Deutsches Elektronen-Synchrotron DESY, Platanenallee 6, 15738 Zeuthen, Germany
}
\affiliation{Institut für Physik, Humboldt-Universität zu Berlin, Newtonstr. 15, 12489 Berlin, Germany}
\author{Simone~Romiti~\orcidlink{0000-0002-6509-447X}}
\email{simone.romiti.1994@gmail.com}
\affiliation{Helmholtz-Institut für Strahlen- und Kernphysik, University of Bonn, Nussallee 14-16, 53115 Bonn, Germany}
\author{Lena~Funcke~\orcidlink{0000-0001-5022-9506}}
\affiliation{Helmholtz-Institut für Strahlen- und Kernphysik, University of Bonn, Nussallee 14-16, 53115 Bonn, Germany}
\affiliation{Transdisciplinary Research Area ``Building Blocks of Matter and Fundamental Interactions'' (TRA Matter), University of Bonn, Bonn, Germany}
\author{Karl~Jansen~\orcidlink{0000-0002-1574-7591}}
\affiliation{
 CQTA, Deutsches Elektronen-Synchrotron DESY, Platanenallee 6, 15738 Zeuthen, Germany
}
\affiliation{
 Computation-Based Science and Technology Research Center, The Cyprus Institute, 20 Kavafi Street,
2121 Nicosia, Cyprus
}
\author{Stefan~K\"uhn~\orcidlink{0000-0001-7693-350X}}
\affiliation{
 CQTA, Deutsches Elektronen-Synchrotron DESY, Platanenallee 6, 15738 Zeuthen, Germany
}
\author{Paolo~Stornati~\orcidlink{0000-0003-4708-9340}}
\affiliation{
ICFO-Institut de Ciencies Fotoniques, The Barcelona Institute of Science and Technology, Mediterranean
Technology Park, Avinguda Carl Friedrich Gauss, 3, 08860 Castelldefels, Barcelona, Spain
}
\author{Carsten~Urbach~\orcidlink{0000-0003-1412-7582}}
\affiliation{Helmholtz-Institut für Strahlen- und Kernphysik, University of Bonn, Nussallee 14-16, 53115 Bonn, Germany}

\date{\today}

\begin{abstract}
In this paper, we examine a compact $U(1)$ lattice gauge theory in $(2+1)$ dimensions and present a strategy for studying the running coupling and extracting the non-perturbative $\Lambda$-parameter.
To this end, we combine Monte Carlo simulations and quantum computing, where the former can be used to determine the numerical value of the lattice spacing $a$, and the latter allows for reaching the perturbative regime at very small values of the bare coupling and, correspondingly, small values of $a$.
The methodology involves a series of sequential steps (i.e., the \textit{step scaling function}) to bridge results from small lattice spacings to non-perturbative large-scale lattice calculations. Focusing on the pure gauge case, we demonstrate that these quantum circuits, adapted to gauge degrees of freedom, are able to capture the relevant physics by studying the expectation value of the plaquette operator, for matching with corresponding Monte Carlo simulations. We also present results for the static potential and static force, which can be related to the renormalized coupling. The procedure outlined in this work can be extended to Abelian and non-Abelian lattice gauge theories with matter fields and might provide a way towards studying lattice quantum chromodynamics utilizing both quantum and classical methods.
\end{abstract}

\maketitle

\section{Introduction\label{sec:intro}}
Quantum field theories are very successful in describing the fundamental laws of nature within the framework of the Standard Model (SM) of particle physics, which unites three of the four known fundamental forces of nature. While many phenomena in the SM can be investigated analytically using perturbation theory, quantum chromodynamics (QCD) is a prominent example of a theory which requires non-perturbative methods in the low-energy regime~\cite{Peskin_1995ev}. This concerns, for instance, the hadron spectrum or the QCD energy scale $\Lambda_\mathrm{QCD}$, which is related to the running coupling of QCD and
is generated entirely dynamically~\cite{ParticleDataGroup:2016lqr}. Therefore, first-principle theoretical calculations of such quantities are of high importance.

The standard approach for non-perturbative computations in quantum field theories is given by the lattice regularization, see e.g. Refs.~\cite{rothe2012lattice,gattringer2009quantum}, in combination with stochastic Monte Carlo (MC) methods based on the Euclidean path integral, pioneered by Wilson~\cite{PhysRevD.10.2445} and Creutz~\cite{Creutz_1982}. 
In this lattice gauge theory (LGT) approach, the theory is regularized by a finite volume and a discretized space-time. In order to make contact with experimental results, the infinite volume and continuum limits need to be taken.
This approach has allowed the computation of many phenomenologically highly relevant quantities, see for instance Ref.~\cite{FlavourLatticeAveragingGroupFLAG:2021npn}, due to significant algorithmic and methodological progress, as well as due to ever-increasing computer power.

Despite these successes, there are still limitations of the MC approach to LGTs.
For example, when the continuum limit is taken, 
 autocorrelation times diverge (sometimes even exponentially fast), see e.g. Refs.~\cite{SCHAEFER201193,schaefer2011algorithms,FlavourLatticeAveragingGroupFLAG:2021npn}.
In this limit of the bare gauge coupling, $g\to0$, the non-perturbative calculation of the running coupling on the lattice could in principle be matched with perturbation theory, even at one loop. This would in turn allow the computation of $\Lambda_\mathrm{QCD}$~\cite{LUSCHER1991221,LUSCHER1993247}. Even though this approach would be natural for this purpose, it is prevented because of the above-mentioned large autocorrelation times of classical MC methods. 
Still, it needs to be pointed out that there are alternative approaches for the non-perturbative computation of the running coupling and hence $\Lambda_\mathrm{QCD}$, see e.g. Refs.~\cite{Fritzsch:2012wq,SOMMER2015155,bruno2017qcd}.

On the other hand, when working with Hamiltonian methods, for instance using quantum computing, there are no autocorrelations.
Hence, such methods offer the potential of following the approach of working in the regime of very small bare couplings, as proposed in Ref.~\cite{PhysRevD.106.114511}. In recent years, the research field of quantum computing has seen much progress, which resulted in various proof-of-concept demonstrations of lattice gauge theory simulations with quantum technologies~\cite{dimeglio2023quantum, banuls2020simulating,Funcke:2023RU}. 

The goal of this work is to develop a framework to compute the running coupling by utilizing both quantum and classical methods,
in order to enable the above-mentioned approach. With present noisy intermediate scale quantum (NISQ) capabilities~\cite{Preskill:2018jim}, it is however impossible to study $(3+1)$-dimensional QCD.
Therefore, in this paper, we aim at a proof-of-concept for this idea in $(2+1)$-dimensional compact $U(1)$ pure gauge theory, with an eventual extension to $(2+1)$-dimensional QED, which shares important properties with $(3+1)$-dimensional QCD, such as confinement and asymptotic freedom. A study of $(2+1)$-dimensional QED has been performed perturbatively, with both two and four-components spinors, in Refs.~\cite{Raviv:2014xna,PhysRevB.96.205113,PhysRevD.94.094013}.
Here, we propose to combine both quantum computing and MC methods and exploit the respective strengths of these two approaches: first, we perform quantum simulations at small values of the bare coupling, to compute the running coupling at small distances. Large volume stochastic simulations~\cite{PhysRevD.106.114511} then allow to determine the lattice spacing by eventually making contact with experimental or phenomenological results, as it is done for Lattice QCD computations~\cite{FlavourLatticeAveragingGroupFLAG:2021npn}.
This work will focus mainly on the aspects of the quantum computing approach and the corresponding numerical techniques. A short description of the MC method, used here, is discussed in Appendix~\ref{app:engap}.

In this paper, we propose a general procedure, based on a step scaling approach, to compute the running coupling as a function of a physical scale, by matching quantum computing and MC techniques in a regime of $g$ where both methods are reliable.
We focus on implementing and testing the feasibility of the method in compact $U(1)$ pure gauge theory as an initial demonstration and propose a follow-up extension to matter fields, which, however, goes beyond the scope of the present work. The inclusion of matter fields will lead to a non-trivial $\beta$-function, rendering the system physically meaningful.
The proposed procedure can be directly generalized to $(2+1)$-dimensional QED but also to non-Abelian lattice gauge theories, and eventually to QCD.

The paper is structured as follows: in Section~\ref{sec:2_1_dim_qed}, we give a concise introduction to the Hamiltonian formulation, the truncation technique for the gauge fields, and the general step scaling method applied to the computation of the running coupling, both in the continuum and on the lattice.
In Section~\ref{sec:Numerical_setup}, we provide a detailed description of the numerical tools used to derive the main results. In particular, we describe the variational quantum circuits developed for the gauge degrees of freedom and the encoding considered.
In Sections~\ref{sec:matching} and~\ref{sec:stepscaling}, we outline the main findings of the paper. Specifically, in Section~\ref{sec:matching} we discuss the results of the expectation value of the plaquette operator for a pure gauge theory on a $3 \times 3$ lattice with periodic boundary conditions, obtained both with a variational quantum algorithm and with exact diagonalization. 
Section~\ref{sec:stepscaling} illustrates the methodology considered for the study of the step scaling method for a pure gauge $3\times 3$ system with open boundary conditions. The quantity analyzed is the static force for two sets of static charge configurations. In Section~\ref{subsec:set1}, we introduce the first static charge configuration and apply the step scaling, starting from the weak coupling regime (Section~\ref{subsubsec:set1w}), using a value of the bare coupling where we have a matching with Monte Carlo (Section~\ref{subsubsec:set1beta}). In Section~\ref{subsec:set2}, the second set of charges is studied.
The method of expressing the static force in terms of a physical scale is presented in Section~\ref{subsec:physscale}.  
In Section~\ref{sec:discussion}, we provide a discussion of the results and give an outlook. Appendix~\ref{app:gray} describes the additional variational quantum circuits developed in this work. Appendix~\ref{app:engap} outlines the Monte Carlo simulations that have been carried out for computing the mass gap, as an eventual alternative for the matching procedure, with a quantitative estimate of the coupling range and a study of finite-size effects.
An extension of the step scaling analysis involving two different basis formulations (electric and magnetic basis) is discussed in Appendix~\ref{app:elmag}, with the corresponding method for data selection.
In Appendix~\ref{app:matter}, we present an in-depth analysis of the $(2+1)$-dimensional QED fermionic Hamiltonian, which will be relevant for future follow-up studies.

\section{\boldmath$(2+1)$-dimensional QED\label{sec:2_1_dim_qed}}
In this section, we present the Hamiltonian describing the compact $U(1)$ lattice QED and discuss the representation of the gauge degrees of freedom for quantum simulations. Additionally, in Section~\ref{subs:runningc}, we introduce the concept of running coupling and illustrate the methodology used in this study to define it.

\subsection{Hamiltonian}
We consider a lattice discretization of the $U(1)$ LGT using Kogut-Susskind staggered fermions~\cite{PhysRevD.11.395,robson1980gauge,ligterink2000many}. 
A naive discretization of the fermionic degrees of freedom leads to the so-called \textit{doubling problem}~\cite{rothe2012lattice,Nielsen:1981hk,PhysRevD.16.3031}, i.e. an incorrect continuum limit of the theory.
In the staggered formulation, the spinor components are distributed on different lattice sites to avoid this problem. The Hamiltonian reads
\begin{equation}\label{eq:fullH}
\hat{H}_\text{tot} = \hat{H}_E + \hat{H}_B + \hat{H}_m + \hat{H}_\text{kin},
\end{equation}
where $\hat{H}_E$ is the electric energy, $\hat{H}_B$ the magnetic energy contribution, $\hat{H}_m$ the fermionic mass term and $\hat{H}_\text{kin}$ the kinetic term for the fermions. 
The electric energy is given by
\begin{eqnarray}\label{eq:hel}
	\hat{H}_E = \frac{g^{2}}{2} \sum_{\vec{r}}\left(\hat{E}^{2}_{\vec{r}, x} 
	+ \hat{E}^{2}_{\vec{r}, y}\right),
\end{eqnarray}
where $\hat{E}_{\vec{r}, \mu}$  is the dimensionless electric field operator that acts on the link emanating from the lattice site with the coordinates $\vec{r}=(r_x,r_y)$ in direction $\mu\in\{x,y\}$. The bare coupling $g$ determines the strength of the interaction, playing a pivotal role throughout the work. 
The second term in $\hat{H}_\text{tot}$, the magnetic interaction, reads
\begin{eqnarray}
\label{eq:hb}
    \hat{H}_B = -\frac{1}{2a^2g^{2}} \sum_{\vec{r}} \left(\hat{P}_{\vec{r}} + \hat{P}_{\vec{r}}^{\dag}
    \right),
\end{eqnarray}
with $a$ the \textit{lattice spacing} and $\hat{P}_{\vec{r}} =  \hat{U}_{\vec{r},x}\hat{U}_{\vec{r}+x,y}\hat{U}^{\dag}_{\vec{r}+y,x}\hat{U}^{\dag}_{\vec{r},y}$ the so-called plaquette operator consisting of a product of the operators $\hat{U}_{\vec{r},x}$ acting on the links of a plaquette of the lattice (with the subscripts notation $\vec{r}+x\equiv (r_x+1,r_y)$ or $\vec{r}+y\equiv (r_x,r_y+1)$). The unitary operators $\hat{U}_{\vec{r},x}$ are related to the discretized vector field $\vec{A}_{\vec{r},\mu}$ as
\begin{equation}
    \hat{U}_{\vec{r},\mu}=e^{iag\vec{A}_{\vec{r},\mu}}.
\end{equation}
They represent the gauge connection between the fermionic fields, and we choose to work with a compact formulation where $ag\vec{A}_{\vec{r},\mu}$ is restricted to $[0,2\pi)$. The lattice vector field is the canonical conjugate variable to the electric field, hence one finds for the commutation relations between $\hat{E}_{\vec{r},\nu}$ and $\hat{U}_{\vec{r'},\mu}$
\begin{align}
    [\hat{E}_{\vec{r},\nu},\hat{U}_{\vec{r'},\mu}]&=\delta_{\vec{r},\vec{r'}}\delta_{\nu,\mu} \hat{U}_{\vec{r},\nu},\label{eq:U_E_commutation_relation1}\\
    [\hat{E}_{\vec{r},\nu},\hat{U}^{\dag}_{\vec{r'},\mu}]&=-\delta_{\vec{r},\vec{r'}}\delta_{\nu,\mu}\hat{U}^{\dag}_{\vec{r'},\nu}.
    \label{eq:U_E_commutation_relation2}
\end{align}
The fermionic mass term is given by 
\begin{eqnarray}\label{eq:hmass}
    \hat{H}_{m} = m \sum_{\vec{r}} (-1)^{r_x+r_y} \hat{\phi}^\dag_{\vec{r}} \hat{\phi}_{\vec{r}},
\end{eqnarray}
where $m$ is the bare lattice fermion mass and $\hat{\phi}_{\vec{r}}$ a and single-component fermionic field residing on site $\vec{r}$, since we start from a continuum formulation with two-component Dirac spinors (see Appendix~\ref{app:matter} for details). The kinetic term corresponds to a correlated fermion hopping between two lattice sites while simultaneously changing the electric field on the link in between\footnote{
Note that here we consider a different kinetic Hamiltonian compared to a previous work~\cite{PhysRevD.106.114511}, by including an additional phase factor and which corresponds to the original Kogut-Susskind formulation.},
\begin{align}   
    \begin{aligned}
        \hat{H}_\text{kin}  =  \frac{i}{2a}&\sum_{\vec{r}}(\hat{\phi}_{\vec{r}}^{\dagger}\hat{U}_{\vec{r},x}\hat{\phi}_{\vec{r}+x} - \text{h.c.}) \\ 
        &-\frac{(-1)^{r_x+r_y}}{2a}\sum_{\vec{r}}(\hat{\phi}_{\vec{r}}^{\dagger}\hat{U}_{\vec{r},y}\hat{\phi}_{\vec{r}+y} + \text{h.c.}).
    \end{aligned}
     \label{eq:hkin}
\end{align}
From now on, we set the $a=1$, unless stated otherwise.
The physically relevant subspace $\mathcal{H}_{\text{ph}}$ of gauge invariant states is given by those that fulfil Gauss's law at each site $\vec{r}$, which reads
\begin{align}
    \begin{aligned}
    \Bigg[\sum\limits_{\mu=x,y}
\left(\hat{E}_{\vec{r}, \mu} -\hat{E}_{\vec{r}- \mu, \mu} \right) - \hat{q}_{\vec{r}} &- Q_{\vec{r}}\Bigg] \ket{\Phi} = 0 \\
&\iff \ket{\Phi} \in \mathcal{H}_{\text{ph}}.
    \end{aligned}
\end{align}
In the above expression, the operators
\begin{equation}
    \hat{q}_{\vec{r}}=\hat{\phi}^\dag_{\vec{r}} \hat{\phi}_{\vec{r}}-\frac{1}{2}\left[1+(-1)^{r_x+r_y+1}\right]
\end{equation}
correspond to the \textit{dynamical charges}, and $Q_{\vec{r}}$ represent \textit{static charges}. The static charges will be particularly relevant for the computation of the static potential in Section~\ref{sec:stepscaling}.
Since in this paper, we are focusing on a $U(1)$ pure gauge theory, we will study only the Hamiltonian $\hat{H}_\text{tot} = \hat{H}_E + \hat{H}_B $. 

We remark that instead of working on the full Hilbert space and enforcing the Gauss's law a posteriori, 
in this work we impose it beforehand and work on a gauge invariant subspace~\cite{Haase2021resourceefficient,PhysRevD.102.094515,PRXQuantum.2.030334,PhysRevD.107.L031503}.

\subsection{Implementation of gauge fields}
The electric field values on a gauge link are unbounded, which leads to infinite dimensional Hilbert space for the gauge degrees of freedom. For a numerical implementation of the Hamiltonian, the gauge degrees of freedom have to be truncated to a finite dimension.
In Ref.~\cite{Haase2021resourceefficient}, the continuous $U(1)$ group is discretized, in the electric basis, to the group of integers $\mathbb{Z}_{2l+1}$, where $l$ introduces a truncation and dictates the dimensionality of the Hilbert space. The discretized gauge fields are constrained to integer values within the range $[-l,l]$, resulting in a total Hilbert space dimension of $(2l+1)^N$, where $N$ denotes the number of gauge fields in the system.
The eigenstates of the electric field operator, $\hat{E}_{\vec{r}, \mu}$, form a basis for the link degrees of freedom (see e.g. Section VI C of Ref.~\cite{RevModPhys.51.659}),
\begin{equation}
    \hat{E}_{\vec{r}, \mu} \ket{e_{\vec{r}, \mu}}=e_{\vec{r}, \mu}\ket{e_{\vec{r}, \mu}} , \ \ \ e_{\vec{r}, \mu} \in [-l,l]
    \, .
\end{equation}
The link operators $\hat{U}_{\vec{r},\mu}$ ($\hat{U}^{\dag}_{\vec{r},\mu}$) act as a raising (lowering) operator on the electric field eigenstates,
\begin{equation}
    \hat{U}_{\vec{r},\mu} \ket{e_{\vec{r}, \mu}}=\ket{e_{\vec{r}, \mu}+1},\ \ \  \hat{U}^{\dag}_{\vec{r},\mu}\ket{e_{\vec{r}, \mu}}=\ket{e_{\vec{r}, \mu}-1}.
\end{equation}
The link operators have the following form~\cite{PRXQuantum.2.030334},
\begin{gather}\label{eq:uoperator}
\hat{U} \mapsto
\begin{bmatrix}
  0 &  \dots & \dots & 0 \\
  1 &  \dots & \dots & 0 \\
  0 &  \ddots & \vdots & 0 \\
  0 &  \dots & 1 & 0\\
\end{bmatrix},\ \ \ \hat{U}^{\dag} \mapsto
\begin{bmatrix}
  0 &  1 & \dots & 0 \\
  0 &  \vdots & \ddots & 0 \\
  0 &  \dots & \dots & 1 \\
  0 &  \dots & \dots & 0\\
\end{bmatrix}.
\end{gather}
With this truncation, unitarity is lost $\hat{U}^{\dag}_{\vec{r},\mu}\hat{U}_{\vec{r},\mu}\neq \mathbb{1}$ but can be recovered in the $l\to \infty$ limit. The commutation relations between the electric field and link operators from Eqs.~\eqref{eq:U_E_commutation_relation1},\eqref{eq:U_E_commutation_relation2} are preserved even for the truncated operators.
Other approaches for the definition of the gauge field operators have been considered in Refs.~\cite{PhysRevD.102.094501,chandrasekharan1997quantum,wiese2013ultracold,Hashizume:2021qbb} and with \textit{qudits}~\cite{meth2023simulating}. 
The errors introduced by finite truncation $l$ have been studied in Refs.~\cite{Kuehn2014,Buyens2017}.

In Appendix~\ref{app:elmag}, we will explore an alternative representation of the Hamiltonian known as the magnetic basis or dual basis, recently explored in Refs.~\cite{PhysRevD.102.114517,Haase2021resourceefficient,PhysRevD.102.094515}. This approach becomes relevant when the coupling constant $g$ decreases and the magnetic term of the Hamiltonian becomes dominant. 
In this regime, the electric basis cannot provide a good approximation of the system with small values of $l$. However, by exploiting the discrete Fourier transform, we can obtain a diagonal expression for the plaquette terms, thus reducing the resources needed for the calculations. With the magnetic formulation introduced in Ref.~\cite{Haase2021resourceefficient}, the group under consideration changes to $\mathbb{Z}_{2L+1}$, where $L$ serves as an additional parameter dictating the discretization. The dimensionality of the Hilbert space remains defined by the truncation parameter $l$. An alternative formulation for the magnetic basis implementation has been explored in Ref.~\cite{PhysRevD.107.L031503}. 
We remark that the Hamiltonian formalism can be extended to non-Abelian gauge groups, like $\mathrm{SU}(2)$ (see e.g. Refs.~\cite{PhysRevD.11.395, PhysRevD.31.3201}). 
For this case, the discretization in the magnetic basis has been investigated more recently in Refs.~\cite{PhysRevD.91.054506, Jakobs:2023lpp,Romiti:2023hbd}. 
Further approaches can be found e.g. in Refs.~\cite{Davoudi:2022xmb,Zache:2023dko,DAndrea:2023qnr}.

Python code implementing the truncation scheme discussed above for the lattice Hamiltonian is available at Ref.~\cite{QEDHamiltrepo}.

\subsection{Running coupling and step scaling}
\label{subs:runningc}
\begin{figure}[htp!]
    \centering
    \includegraphics[width=1.0\columnwidth]{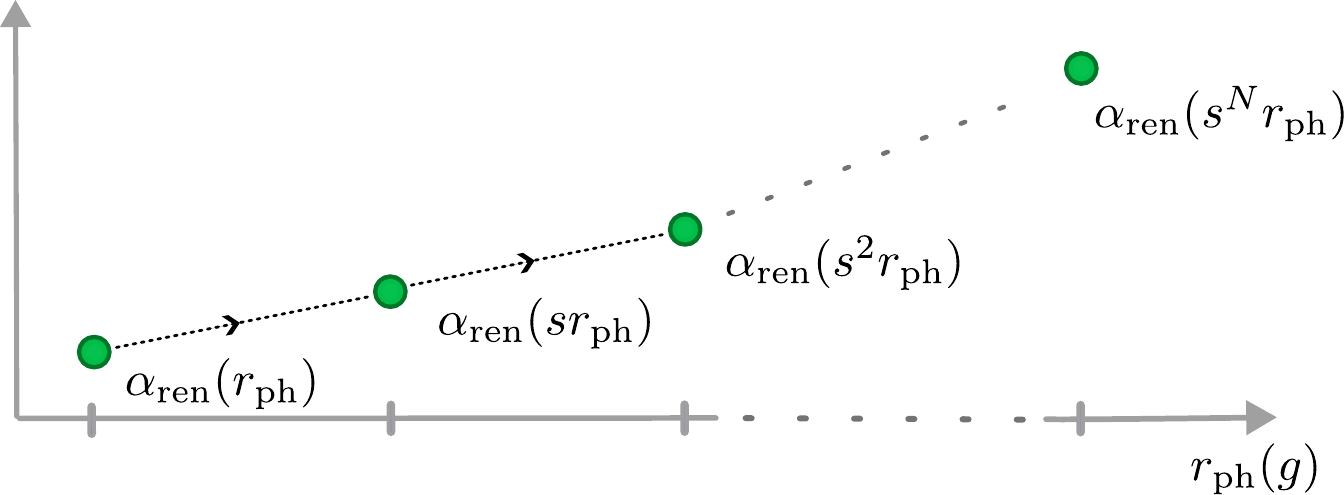}
    \caption{\textbf{Step scaling in the continuum theory:} Schematic illustration of the method (See text for more details).}
    \label{fig:step_cont}
\end{figure}

\begin{figure}[htp!]
    \centering
    \includegraphics[width=0.45\textwidth]{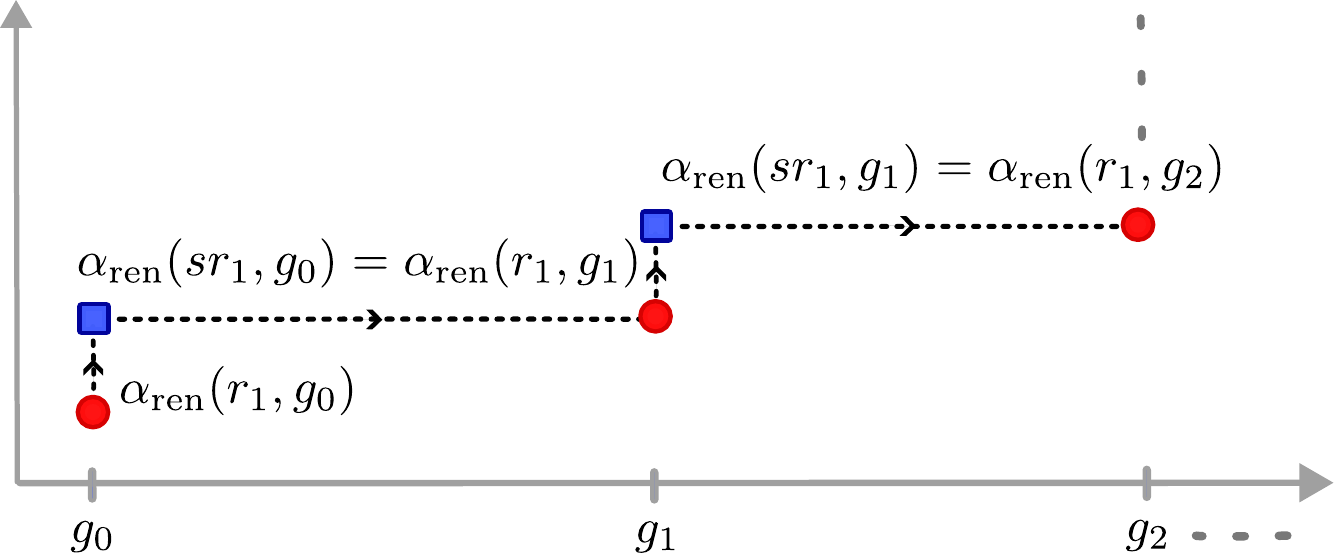}
    \caption{\textbf{Step scaling on the lattice:} Schematic illustration of the method (See text for more details).}
    \label{fig:step_latt}
\end{figure}

The \textit{step scaling} approach is a computational method employed for the determination of the running coupling, introduced in Ref.~\cite{LUSCHER1991221} and used also for instance in Refs.~\cite{Luscher:1993gh,DellaMorte:2004bc}. For a general description based on the Schr\"odinger functional approach see Refs.~\cite{Luscher:1991wu,lüscher1998advanced}. 
Let us assume we define a running, renormalized coupling $\alpha_\mathrm{ren}(r_\text{ph})$ at a physical scale $r_\text{ph}(g)$. We then define the step scaling function $\sigma_s$ in the continuum from
\begin{equation}\label{eq:stepscalingfunc}
  \sigma_s(\alpha_\mathrm{ren}(r_\text{ph})) = \alpha_\mathrm{ren}(sr_\text{ph})\,,\qquad s\in\mathbb{R}^+,
\end{equation}
which can be understood as an integrated form of the $\beta$-function of the theory. 
Starting from $\alpha_\mathrm{ren}(r_{\text{ph}})$, we then apply the function in Eq.~\eqref{eq:stepscalingfunc}. 
The step then is repeated, going to $\alpha_\mathrm{ren}(s^2r_{\text{ph}})$ and subsequent values, by creating the steps in Fig.~\ref{fig:step_cont}.
This method can be iterated up to arbitrary $N+1$ steps, obtaining $\alpha_\mathrm{ren}(s^Nr_{1,\text{ph}})$ and thus getting the running coupling as a function of the physical scale. 
The goal of this work is to compute the step scaling function non-perturbatively on the lattice, by starting with some distance $r$ in lattice units and with a bare coupling $g$.
The lattice spacing is encoded in the coupling which is an implicit function of $a$ in physical units.
We fix two scales, $r_1$ and $r_2\equiv s\cdot r_1$ in lattice units and compute the renormalized coupling at a fixed value of $g$, i.e. $\alpha_\mathrm{ren}(r_1, g_0)$ and $\alpha_\mathrm{ren}(sr_1, g_0)$, which corresponds to $\sigma_s(\alpha_\mathrm{ren}(r_1, g_0))$ on the lattice. We tune $g$, finding the value where $\alpha_\mathrm{ren}(r_1, g_1)=\alpha_\mathrm{ren}(sr_1, g_0) $. 
The corresponding sequence of steps can be illustrated in Fig.~\ref{fig:step_latt}.

In this paper, we consider a lattice calculation and then we convert the lattice distances into physical ones with an artificial value of $a$ in physical units and $r_{\text{ph}}=ar$. Where, as mentioned in the introduction, the numerical value of the lattice spacing can be obtained in principle with large-volume Monte Carlo computations. Since $a$ in physical units is a function of the bare coupling, by decreasing $g$, we change the physical distance to smaller values. In this way, we get the running of the coupling as a function of the physical scale.
Results of the application of this method are discussed in Section~\ref{subsec:physscale}.
We use the static force, $F(r,g)$, as the physical quantity of interest, focusing in particular on the dimensionless quantity $r^2 F(r,g)$, with $g$ the bare coupling at which the force is computed and $r$ the distance between two static charges. 
The calculation of the static force involves the application of a discrete derivative, approximated as $\frac{\partial V}{\partial r} \simeq \frac{V(r_2)-V(r_1)}{r_2 - r_1}$, where $V(r_i)$ denotes the static potential between two static charges separated by $r_i$. 
This potential is, for short distances, proportional to a logarithmic Coulomb term $V(r)\sim \alpha_{\text{ren}}\log r$~\cite{Loan:2002ej} on the lattice. Thus $r^2 F(r,g)$ can be related eventually to the renormalized coupling.
For the analysis of the step scaling, we need two values of the static force,
\begin{subequations}
\begin{align}
    F(r_2,g) &=\frac{V(r_3,g)-V(r_2,g)}{r_3-r_2}, \\
    F(r_1,g) &=\frac{V(r_2,g)-V(r_1,g)}{r_2-r_1}.
\end{align}
\end{subequations}
Therefore, it is necessary to involve three distances, namely $r_1$, $r_2$ and $r_3$, in the calculation of the step scaling function. Note that in this paper we introduce the step scaling method for a pure gauge $U(1)$ theory, once we will include matter fields we will have a non-trivial running coupling.

\section{Numerical setup\label{sec:Numerical_setup}}
After discretizing and truncating the $U(1)$ gauge group to the discrete group $\mathbb{Z}_{2l+1}$, the gauge fields can be represented in the electric basis as
\begin{subequations}\label{eq:EUdef}
\begin{align}
    \hat{E} &= \sum_{i=-l}^l i \ket{i}_{\text{ph}}\bra{i}_{\text{ph}}, \\
    \hat{U} &= \sum_{i=-l}^{l-1} \ket{i+1}_{\text{ph}}\bra{i}_{\text{ph}},\\
    \hat{U^\dagger} &= \sum_{i=-l+1}^l \ket{i-1}_{\text{ph}}\bra{i}_{\text{ph}}.
\end{align}
\end{subequations}
For numerical calculations, it is advantageous to employ a suitable encoding that accurately represents the physical values of the gauge fields.
In this work, we consider the \textit{Gray encoding} (see, e.g., Ref.~\cite{PhysRevA.103.042405}). With this approach, the minimum number of qubits required per gauge variable is $q_\text{min} = \lceil \log_2(2l+1) \rceil.$ Thus, it will be convenient for the implementation on a quantum circuit to consider a subset of truncation values ($l=1,3,7,15,...$), which allows only a single state to be excluded with the same amount of resources. For instance, three qubits are required for both $l=2$ and $l=3$. However, with the former, only five configurations are considered physical, whereas with the latter, we can include seven physical states.

The state of a qubit can be defined as a vector in a 2-dimensional complex vector space $\mathbb{C}^2$, with
$\ket{0}=( 1, 0)^t$ and $\ket{1}=(0, 1)^t$
as the computational basis~\cite{nielsen2010quantum}.
The quantum operations, or \textit{gates}, on a single qubit can be described by $2\times 2$ unitary matrices. Thus, for numerical implementations, we express the Hamiltonian in terms of a sum of Pauli matrices. One can also employ a grouping strategy to identify subsets of Pauli strings present in the Hamiltonian, thereby reducing the necessity for independent circuit evaluations~\cite{tilly2022variational}.
In the following, we also adopt the convention that the least significant qubit (designated by the zero index, $q_0$) occupies the rightmost position in the tensor product, as illustrated by $\ket{q_1q_0}$ and $\bra{q_1q_0}$. 
Let us now consider, as an example, the case of smallest truncation $l=1$, where we have the three physical states $\ket{j}_{\text{ph}}$
for $j\in \{-1,0,1\}$.
These states can be encoded using only two qubits in a Gray code way, as shown in the following equations:
\begin{subequations}
\begin{align}
   \ket{-1}_{\text{ph}} \mapsto &~\ket{00}, \\
   \ket{0}_{\text{ph}} \mapsto &~\ket{01}, \\
   \ket{1}_{\text{ph}} \mapsto &~\ket{11},
\end{align}
\end{subequations}
we then call the state $\ket{10}$ ``unphysical'', since it is outside of this truncated Hilbert space. 
The expressions for the electric field and link operators then become
\begin{subequations}
\begin{align}
    \hat{E} &\mapsto -\ket{00}\bra{00} +\ket{11}\bra{11}, \\
    \hat{U} &\mapsto  \ket{01}\bra{00} +\ket{11}\bra{01}, \\
    \hat{U^\dagger} &\mapsto  \ket{00}\bra{01} +\ket{01}\bra{11} .
\end{align}
\end{subequations}
In this study, we adopt a variational approach to determine the physical quantities of interest. Specifically, we employ the Variational Quantum Eigensolver (VQE) method~\cite{peruzzo2014variational} that aims to find the ground state of a given Hamiltonian. 
Executing a VQE algorithm requires an input quantum circuit with parametrized gates, called Ansatz circuit, and a classical optimizer. The optimization starts with an initial set of values for the gate parameters, that can be randomly chosen, and will be optimized in the execution. In the rest of the paper, we consider a set of parameters, where the probability of being in a vacuum state (i.e. $\ket{0}_{\text{ph}}$) is non-zero for every gauge field.
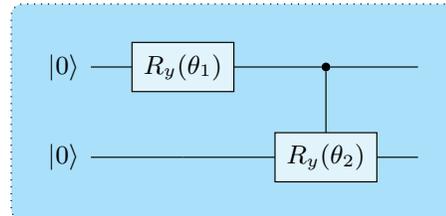
\begin{figure}[htp!]
\centering
\tikzset{operator/.append style={fill=cyan!10}}  
\noindent   \resizebox{0.7\columnwidth}{!}{
 \begin{quantikz}[thin lines] 
 \gategroup[2,steps=5,style={dotted,rounded corners,fill=cyan!30,inner sep=10pt},background]{} 
& \lstick{$\ket{0}$} &\gate{R_y(\theta_1)} & \ctrl{1} & \qw\\
& \lstick{$\ket{0}$} & \qw & \gate{R_y(\theta_2)} & \qw     
\end{quantikz}
}   
\caption{\textbf{Variational circuit for Gray encoding with $\bm{l=1}$:} \textit{Vacuum state} is $\ket{01}$, and state $\ket{10}$ is excluded.}
\label{grayl1}
\end{figure}
The essence of the approach considered here is to exclude unphysical states directly within the quantum circuit. This is achieved by implementing a customized set of parameterized quantum gates designed to produce the correct final combination of states. 
With this method, we aim to efficiently identify the desired physical results while reducing the computational overhead\footnote{We also considered keeping the state $\ket{10}$ as a higher physical state $\ket{2}_{\text{ph}}$ and use a generic variational Ansatz. However, the VQE results did not have a high fidelity. Therefore, we will not describe this option further. It may be considered in future work.}. 
For the truncation $l=1$, we can use the circuit in Fig.~\ref{grayl1} to represent a gauge field. The action of the circuit is straightforward: starting from the state $\ket{00}$, setting both parameters $\theta_1$ and $\theta_2$ to zero allows for the exploration of the physical state $\ket{-1}_{\text{ph}}$. The introduction of a non-zero value for $\theta_1$ allows the state to change to $\ket{01}$, which represents the  \textit{vacuum state} $\ket{0}_{\text{ph}}$, with a certain probability. A complete rotation occurs if $\theta_1=\pi$, resulting in the exclusive presence of the second state with a probability of 1.0. Subsequently, the second controlled gate operates only when the first qubit is $\ket{1}$, limiting the exploration to $\ket{11}$ (i.e., $\ket{1}_{\text{ph}}$) and excluding $\ket{10}$.

This procedure can be expanded to arbitrary $l$, allowing the exclusion of unphysical combinations, and to multiple gauge fields with entangling gates. For further details, refer to Appendix~\ref{app:gray} which provides an extension to three additional values of truncation.

\section{Results: Matching strategy for \boldmath$3\times 3$ PBC system}
\label{sec:matching}
\begin{figure}[htp!]
    \centering
    \includegraphics[width=0.3\textwidth]{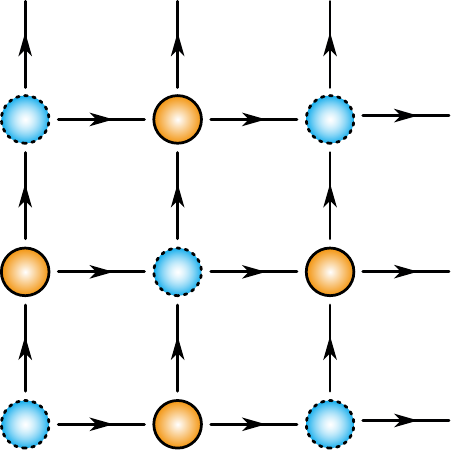}
    \caption{\textbf{Illustration of a $\bm{3\times3}$ lattice with periodic boundary conditions:} The spheres represent the matter sites, where blue dashed (orange solid) circles indicate sites with even (odd) parity. The lines connecting the vertices represent the gauge links where the arrows indicate the orientation of the lattice. The links sticking out on top (to the right) indicate the periodic boundary conditions and connect to the vertices on the bottom (left).}
    \label{fig:3x3pbc}
\end{figure}

\begin{figure}[htp!]
    \centering
    \includegraphics[width=1.0\columnwidth]{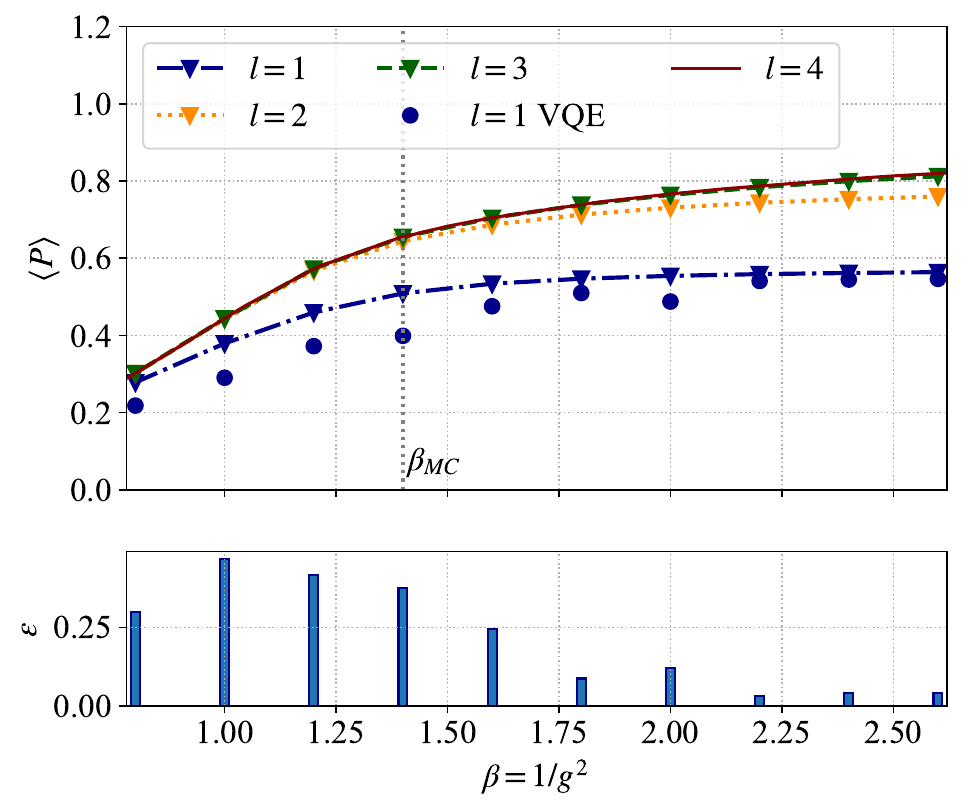}
    \caption{(\textit{top}) \textbf{Plaquette expectation value for $\bm{3\times 3}$ pure gauge PBC system:} exact diagonalization results with truncation $l \in [1,4]$ (\textit{lines with low triangles} or \textit{solid line}) and VQE results (noise-free simulations with infinite shots) with  $l=1$ (\textit{circles}). The \textit{dotted} vertical line corresponds to $\beta$ matching with MC. (\textit{bottom}) \textbf{Relative error $\bm{\epsilon}$:} Comparison between VQE data and exact results. The error depends on the convergence of the optimization reaching a given tolerance and the initial set of parameter values in the quantum gates.}
    \label{plaqpbcvqe}
\end{figure}

In this section, we describe the matching between the VQE approach and Markov Chain Monte Carlo (MCMC). For our study, we consider the pure gauge case, i.e.\ the theory without fermionic fields, on a $3\times3$ lattice with periodic boundary conditions (PBC) (see Fig.~\ref{fig:3x3pbc} for an illustration). The quantity we analyze is the expectation value of the plaquette operator,
\begin{equation}
    \expectationvalue{P} = \expectationvalue{\sum_{\vec{r}} \left(\frac{\hat{P}_{\vec{r}} + \hat{P}_{\vec{r}}^\dagger}{2}\right)}.
\end{equation}
We focus on bare couplings in the interval corresponding to $0.8\leq\beta=1/g^2\leq 2.6$, selected to be in a regime accessible with MCMC methods. We analyze the convergence behavior of the results with exact diagonalization (ED) with respect to the truncation parameter $l$, as illustrated by the lines in the upper panel of Fig.~\ref{plaqpbcvqe}. 
The results from ED show that with increasing values of $l$ we observe convergence, and for the range of couplings chosen here $l=3$ is sufficient to reliably determine the plaquette expectation value, as the results for $l=3$ and $l=4$ are essentially indistinguishable. For a truncation $l=2$ we see slight deviations for larger values of $\beta$. While for $l=1$ deviations towards larger values of $\beta$ are noticeable, the data still qualitatively reproduce the behavior observed for larger values of $l$.
For the VQE approach, we have developed a quantum circuit with an entanglement structure connecting all gauge fields. 
The resources required for running the circuit are shown in Table~\ref{tab:res3x3pbc}. 
\begin{table}[htp!]
\centering
\begin{tabular}{|p{0.4cm}||p{1.4cm}|p{1.5cm}|p{2cm}|p{2.1cm}| }
 \hline
 \multicolumn{5}{|c|}{Resources Estimation $3\times3$ PBC system} \\
 \hline
 $l$ & $\#$ Qubits & $\#$ CNOTs & CNOT Depth & $\#$ Parameters\\
 \hline
 1 & 20  &  1280   & 1152 &  200 \\
 3 & 30  & 2200    & 1748 &  445 \\
 \hline
\end{tabular}
\caption{\textbf{Resources required for the variational circuit for Gray encoding:} In a pure gauge $3\times 3$ PBC system the ten dynamical gauge fields can be simulated with the specified total number of qubits. Additionally, we quantify the total count of CNOT gates and the CNOT depth, representing the layers of CNOT gates in the circuit. The rightmost column displays the total number of parameters in the variational Ansatz.}
\label{tab:res3x3pbc}
\end{table}
Due to the limited resources available on current quantum devices, we focus on truncation $l=1$ for a proof of principle.
In order to benchmark the approach, we classically simulate the VQE, assuming a noise-free quantum device.
For the optimization, we employ the SLSQP~\cite{kraft1988software} classical optimizer and infinite shots, i.e.\ number of measurements, setting as the first goal only the expressivity of the quantum circuit.
After applying Gauss's law only 10 of the 18 links remain dynamical, thus we need 20 qubits for the computation. 
As illustrated in Fig.~\ref{plaqpbcvqe}, the top panel shows the VQE results for this truncation, indicated by \textit{circles}, along with the relative error with respect to the exact values in the bottom panel. 
These results are in line with the plaquette curve from ED. 

For the future, a closer examination of the entanglement structure could help to improve the accuracy of the data and optimize the scalability of the gate number.
Such exploration should focus on three main purposes: to improve our understanding of the interplay between circuit and lattice structure, to extend the results to higher truncation while preserving the depths of the circuits and to prepare for analysis on quantum hardware platforms.

The vertical line in Fig.~\ref{plaqpbcvqe} marks the point where the exact results for the plaquette in the Hamiltonian formalism align with those obtained by the Monte Carlo approach.
For the MC simulations, part of the authors are investigating the same $U(1)$ gauge theory in $(2+1)$ dimensions in the Lagrangian formalism and take the continuum limit in time, keeping the spatial lattice spacing fixed. A first, preliminary account of this work can be found in Ref.~\cite{Funcke:2022opx}.
This procedure returns a $\beta_{MC}$-value for which we know the corresponding bare coupling value of the discretized theory. Thus, the spatial lattice spacing is identical up to lattice artefacts. Since this analysis is still ongoing, we use here the preliminary value $\beta_{MC}= 1.4$.
At this value of the coupling, we can then perform large-volume Monte Carlo simulations and set the physical scale.

\section{Results: Step scaling approach}
\label{sec:stepscaling}
\begin{figure*}
    \centering
    \includegraphics[width=0.95\textwidth]{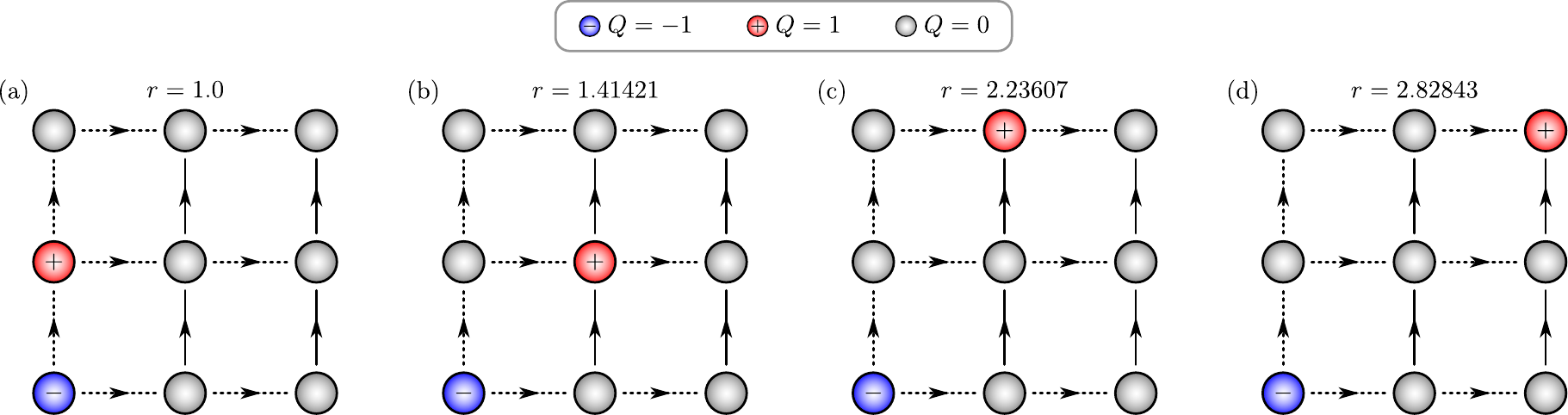}
    \caption{\textbf{Illustration of the static charge configurations:} \textit{blue} spheres with a $-$ (\textit{red} spheres with a $+$) correspond to sites carrying a negative (positive) static charge, grey spheres to sites where no static charge is present. The solid arrows indicate the dynamical links, and the dashed arrows are the nondynamical ones. The different panels correspond to different distances of the charges with $r=1.0$ (a), $r=\sqrt{2}=1.414$ (b), $r=\sqrt{5}=2.236$ (c), and $r=\sqrt{8}=2.828$ (d).}
    \label{fig:staticchconfig}
\end{figure*}
In this section, we illustrate the methodology for the pure gauge case on a $3\times 3$ lattice with Open Boundary Conditions (OBC). Two static charges of opposite values are placed on two sites, as in Fig.~\ref{fig:staticchconfig}. The choice of boundary conditions allows us to obtain more distinct lattice distances than the periodic case. We focus on two sets of distances to generalize any findings regarding the coupling behavior. We have tested all the combinations of the five possible distances for two static charges on a pure gauge lattice, $C(5,3)=\frac{5!}{3!(5-3)!}$, and have chosen the two combinations with more points in the step scaling procedure below a certain threshold for the bare coupling, i.e. $\beta\leq 10^2$.
Note that with a system size of $3\times 3$ sites and a range of $\beta$s considered here, we always work at distances below the confinement scale, and the correlation length is given by the small system size~\cite{lüscher1998advanced}.

In the following analysis, a variational quantum algorithm is used to calculate the static potential at different distances, and the results are compared with those derived from exact diagonalization.
The data presented here are computed with a combination of two classical optimizers: we performed a first minimization with NFT~\cite{PhysRevResearch.2.043158}, which gave us fidelities of up to $\sim 95\%$ (noise-free simulations with $\sim \mathcal{O}(10^4)$  shots). As the coupling decreases, higher precision in the VQE results becomes necessary. Consequently, we have used the final optimal parameters as a starting point for a new optimization with COBYLA~\cite{powell1994direct} and a larger number of shots ($\sim \mathcal{O}(10^6))$. This aspect is crucial for our objectives, as the values of the static forces in the weak coupling regime are almost equivalent. 
A better understanding of the entanglement structure may be helpful for increasing the precision with fewer shots. We will consider a more in-depth analysis of this in future work.
In Table~\ref{tab:res3x3obc}, we show the resource estimation for three values of the truncation parameter $l$. 
\begin{table}[H]
\centering
\begin{tabular}{ |p{0.4cm}||p{1.4cm}|p{1.5cm}|p{2cm}|p{2.1cm}|  }
 \hline
 \multicolumn{5}{|c|}{Resource Estimation $3\times 3$ OBC system} \\
 \hline
 $l$ & $\#$ Qubits & $\#$ CNOTs & CNOT Depth & $\#$ Parameters\\
 \hline
 1 & 8  &  176   & 168 & 32  \\
 3 & 12  &  304   & 260 &  70 \\
 7 & 16  &  456   & 354 &  120 \\
 \hline
\end{tabular}
\caption{\textbf{Resources required for the variational circuit for Gray encoding:} In a pure gauge $3\times 3$ OBC system, the four dynamical gauge fields can be simulated with the specified total number of qubits. Additionally, we quantify the total count of CNOT gates and the CNOT depth, representing the layers of CNOT gates in the circuit. The rightmost column displays the total number of parameters in the variational Ansatz.}
\label{tab:res3x3obc}
\end{table}

\subsection{Step scaling results for static forces \boldmath$F(r_1=1,g)$ and \boldmath$F(r_2=\sqrt{5},g)$}\label{subsec:set1}
In this section, we focus on illustrating the step scaling method using a variational approach and exact diagonalization as a reference, for the configuration with charges placed as in Fig.~\ref{fig:staticchconfig} (panel (a), (c) and (d)), and relative distances on the lattice $r_1=1$, $r_2=\sqrt{5}$ and  $r_3=\sqrt{8}$, respectively.

\subsubsection{Illustration of step scaling from $\beta=10^2$}
\label{subsubsec:set1w}
\begin{figure}[htp!]
    \centering
    \includegraphics[width=1.0\columnwidth]{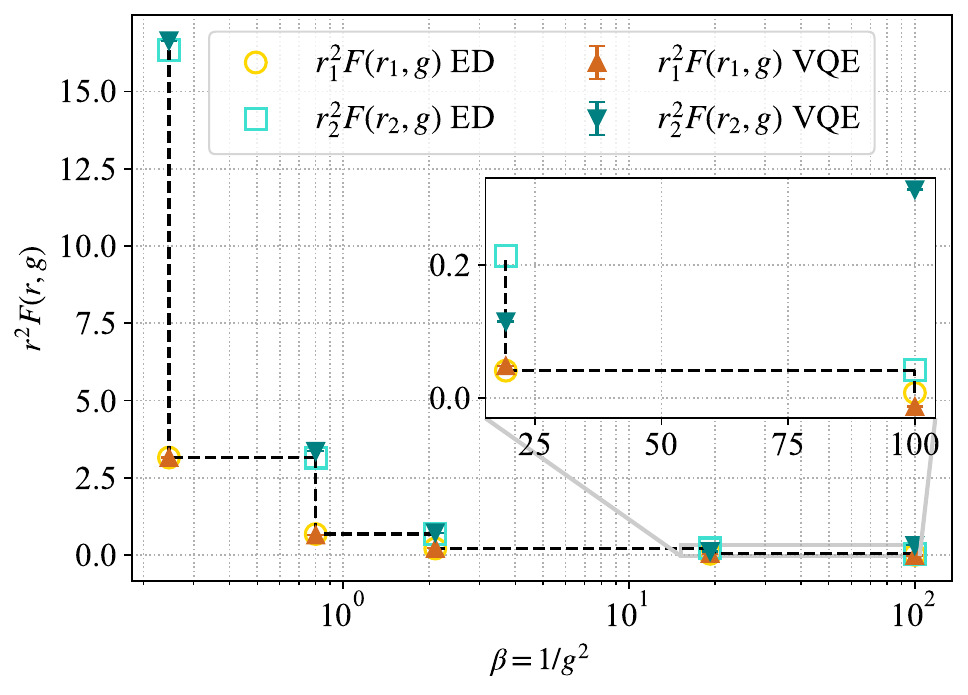}
    \caption{\textbf{Step scaling results for static forces $\bm{F(r_1=1,g)}$ and $\bm{F(r_2=\sqrt{5},g)}$, electric basis and $\bm{l=1}$:}  Here we used the smallest truncation to illustrate the method.
    From the weak coupling regime $\beta=10^2$, the static forces are computed following a steps procedure, both with ED and VQE (noise-free simulations with shots).
    ED results for $r_1^2F(r_1=1,g)$ ($r_2^2F(r_2=\sqrt{5},g)$) are displayed with \textit{circles} (\textit{squares}) and corresponding VQE results with \textit{up(down)ward-pointing triangles}.  In the simulations, a combination of NFT and COBYLA optimizer was considered and a finite number of shots defines the error bars, which are smaller than the markers.}
    \label{step_g0.1}
\end{figure}

We show the step scaling procedure starting from a weak coupling regime. As explained in Section~\ref{subs:runningc}, we repeat the steps until we reach a certain value of the bare coupling.
In Fig.~\ref{step_g0.1}, we follow the step scaling procedure starting from $\beta=10^2$ and moving to the left with a variational Ansatz (\textit{up(down)ward-pointing triangles}) and exact results (\textit{empty circles/squares}).
One can see that the precision required for small couplings increases, because of the small values required in the step scaling function. Considering only the electric basis becomes more difficult, as the superposition within the ground state expands significantly towards weaker couplings. It is thus advisable to start at large $\beta$-values with the magnetic basis and monitor the convergence of results throughout the process towards smaller $\beta$-values.

\subsubsection{Start from $\beta =1.4$ to perturbative regime }
\label{subsubsec:set1beta}
\begin{figure}[htp!]
    \centering
    \includegraphics[width=1.0\columnwidth]{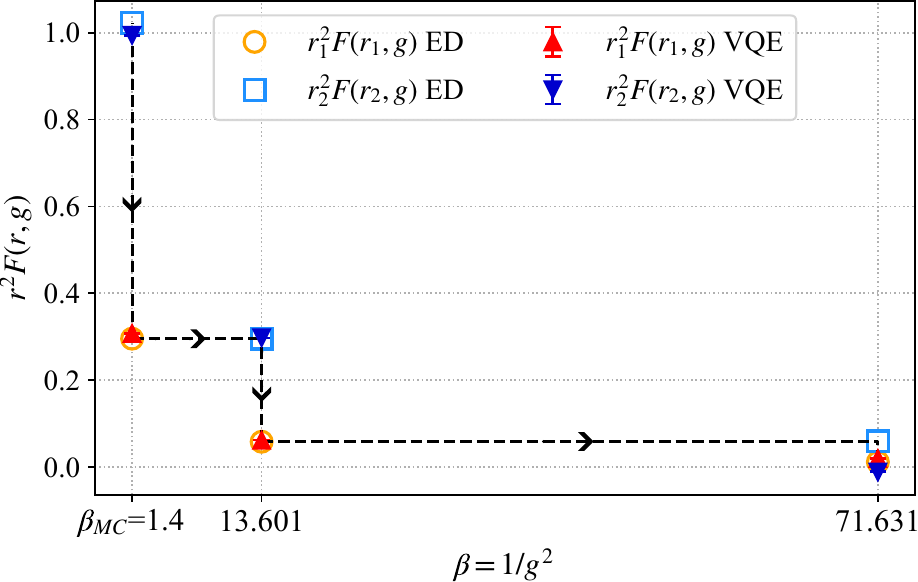}
    \caption{\textbf{Step scaling results for static forces $\bm{F(r_1=1,g)}$ and $\bm{F(r_2=\sqrt{5},g)}$, electric basis and $\bm{l=1}$:}  From $\beta_{MC}=1.4$ and in a range of couplings within $\beta\leq 10^2$, the static forces are computed following a steps procedure, both with ED and VQE (noise-free simulations with shots).
    ED results for $r_1^2F(r_1=1,g)$ ($r_2^2F(r_2=\sqrt{5},g)$) displayed with \textit{circles} (\textit{squares}) and corresponding variational results with \textit{up(down)ward-pointing triangles}. In the simulations, a sequential combination of two optimizers NFT and COBYLA was considered and a finite number of shots defines the error bars, which are smaller than the markers.}
    \label{fig:stepl1light}
\end{figure}

Here, we discuss the variational results for the step scaling method, starting from the value of the bare coupling where we have a matching with MC, see Section~\ref{sec:matching}, and continuing towards a weaker regime. We first illustrate the procedure with a fixed truncation, $l=1$, and then discuss higher truncations, involving also a magnetic representation.
Starting from $\beta_{MC}=1.4$, we compute $r_1^2F(r_1,g_0)$ and $r_2^2F(r_2,g_0)$. Next, using the result of the static force at a distance $r_1$, the bare coupling $g$ is adjusted to a reduced value until a new $r_2^2F(r_2,g_1)$ is obtained. This step scaling process is then repeated using a similar approach as in the previous paragraph, but aimed at the weak coupling regime. The comparison between variational results (denoted by \textit{up/downward-pointing triangles}) and exact results (denoted by \textit{empty circles/squares}) is illustrated in Fig.~\ref{fig:stepl1light}.

\begin{figure}[htp!]
    \centering
    \includegraphics[width=1.0\columnwidth]{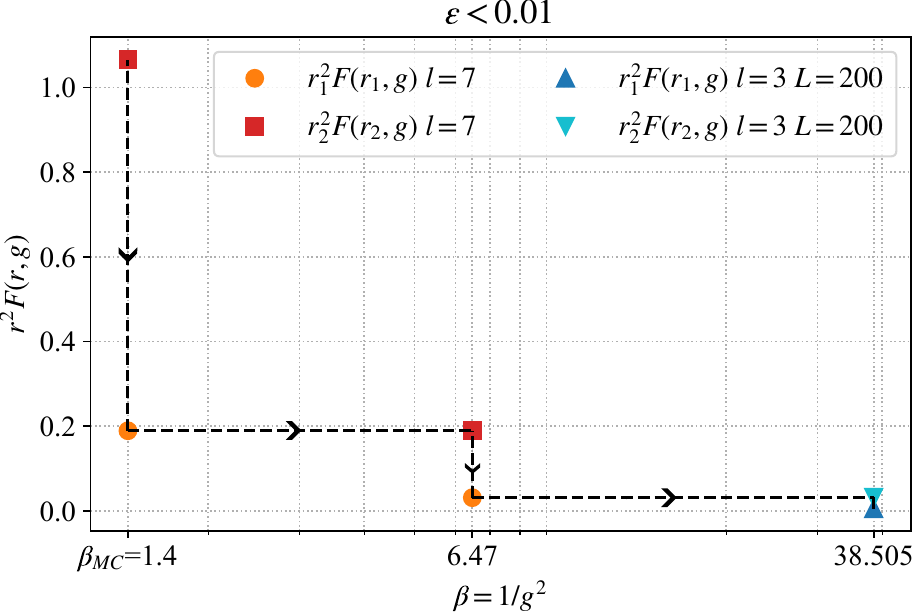}
    \caption{\textbf{Step scaling results for static forces $\bm{F(r_1=1,g)}$ and $\bm{F(r_2=\sqrt{5},g)}$, electric (magnetic) basis and $\bm{l=7}$ ($\bm{l=3,L=200}$), with relative error $\bm{\epsilon<0.01}$:} 
    In contrast to Fig.~\ref{fig:stepl1light}, we consider here a higher truncation and show
    ED results with electric basis for $r_1^2F(r_1=1,g)$ ($r_2^2F(r_2=\sqrt{5},g)$) with truncation value $l=7$ displayed with \textit{circles}(\textit{squares}) and with magnetic basis for $r_1^2F(r_1=1,g)$ ($r_2^2F(r_2=\sqrt{5},g)$) with $l=3$ and discretization values $L=200$ displayed with \textit{up(down)ward-pointing triangles}.}
    \label{step_l7lightblue2}
\end{figure}

To give physical meaning to the data, it is essential to increase the truncation parameter $l$ applied to the gauge operators until independent solutions are obtained.
Given the limited resources, we adopt a strategy involving an interplay between electric and magnetic basis. Starting from the electric formulation, we progressively increase $l$ until convergence is achieved within the desired range, $1.4\leq \beta \leq 10^2$. Once a reference value is established, we restrict ourselves to $l=7$, which will result in a total of 16 qubits, which seems feasible on current quantum hardware.
Next, following the procedure described in Appendix~\ref{app:elmag}, we find the value of the bare coupling where the accuracy of the electric basis is not sufficient anymore (i.e.\ exceeding a relative error of $\epsilon \geq 0.01$). At this point, we move on to the magnetic basis with $l=3$ and discretization $L=200$, parameters that give us reliable results. Initially, we conducted tests with $l=7$ also for the magnetic basis, maintaining an equal number of qubits for each register as for the electric one. The outcomes proved to be comparable to those obtained with $l=3$. Consequently, we can decrease the computational resources required while preserving a high level of accuracy in the solutions.
Fig.~\ref{step_l7lightblue2} illustrates the step-scaling method employing the technique described above. Similarly, we proceed through the weak coupling regime by increasing $\beta$ and constructing the steps accordingly. The VQE approach is still ongoing, with the study of new entanglement structures.


\subsection{Step scaling results for static forces \boldmath$F(r_1=\sqrt{2},g)$ and \boldmath$F(r_2=\sqrt{5},g)$}\label{subsec:set2}
In this section, the analysis is repeated for a new set of distances,  $r_1=\sqrt{2}$, $r_2=\sqrt{5}$ and $r_3=\sqrt{8}$, Fig.~\ref{fig:staticchconfig} (panel (b), (c) and (d)). Here, we solely explore the step scaling starting from $\beta_{MC}$. The results are then combined in Section~\ref{subsec:physscale} with the previous set of distances, in order to show the dependence of $r^2 F(r,g)$ in terms of a physical scale.

\subsubsection{Start from $\beta=1.4$ to perturbative regime}
\label{subsubsec:set2beta}
\begin{figure}[htp!]
    \centering
    \includegraphics[width=1.0\columnwidth]{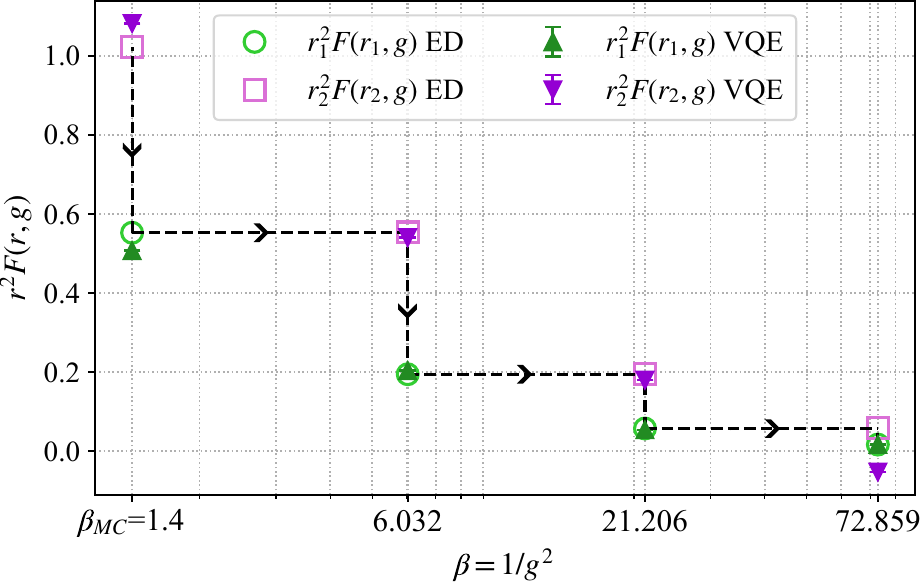}
    \caption{\textbf{Step scaling results for static forces $\bm{F(r_1=\sqrt{2},g)}$ and $\bm{F(r_2=\sqrt{5},g)}$, electric basis and $\bm{l=1}$:}  Similar approach as in Fig.~\ref{fig:stepl1light}, with a different set of distances. From $\beta_{MC}=1.4$ and in a range of couplings within $\beta\leq10^2$, the static forces are computed following a steps procedure, both with ED and VQE (noise-free simulations with shots).
    ED results for $r_1^2F(r_1=\sqrt{2},g)$ ($r_2^2F(r_2=\sqrt{5},g)$) displayed with \textit{circles} (\textit{squares}) and corresponding variational results with \textit{up(down)ward-pointing triangles}. In the simulations, a sequential combination of two optimizers NFT and COBYLA was considered and a finite number of shots defines the error bars, which are smaller than the markers.}
    \label{fig:stepl1purple}
\end{figure}

The step scaling procedure is illustrated in Fig.~\ref{fig:stepl1purple} in the fixed bare coupling interval. In this case, four steps are observed within the range $1.4\leq \beta \leq 10^2$.

\begin{figure}[H]
    \centering
    \includegraphics[width=1.0\columnwidth]{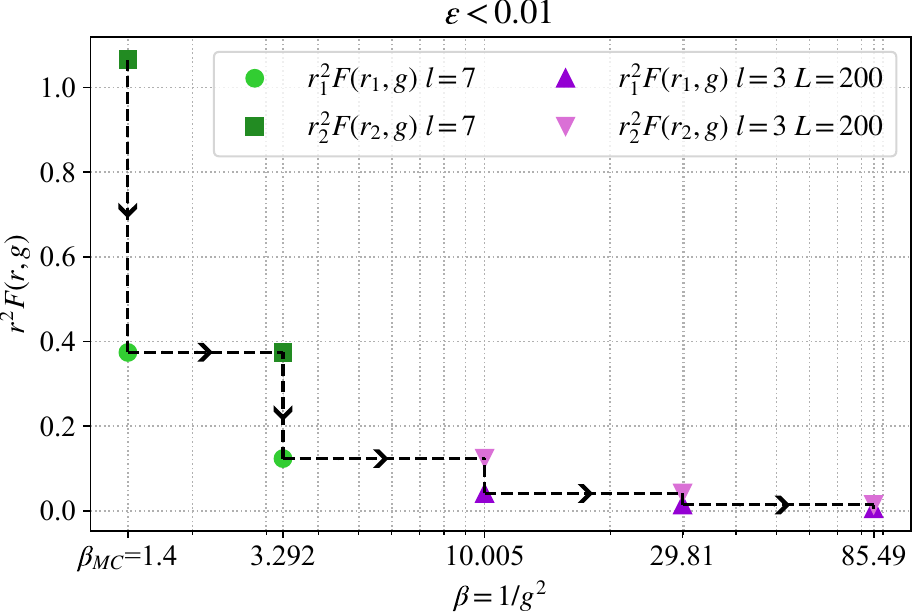}
    \caption{\textbf{Step scaling results for static forces $\bm{F(r_1=\sqrt{2},g)}$ and $\bm{F(r_2=\sqrt{5},g)}$, electric (magnetic) basis and $\bm{l=7}$ ($\bm{l=3,L=200}$), with relative error $\bm{\epsilon<0.01}$:} In contrast to Fig.~\ref{fig:stepl1purple}, we consider here a higher truncation and show
    ED results with electric basis for $r_1^2F(r_1=\sqrt{2},g)$ ($r_2^2F(r_2=\sqrt{5},g)$) with truncation value $l=7$ displayed with \textit{circles}(\textit{squares}) and with magnetic basis for $r_1^2F(r_1=\sqrt{2},g)$ ($r_2^2F(r_2=\sqrt{5},g)$) with $l=3$ and discretization values $L=200$ displayed with \textit{up(down)ward-pointing triangles}}
    \label{step_l7purple2}
\end{figure}

We apply the same technique with higher truncations for this set of distances, Fig.~\ref{step_l7purple2}. Also in such a case, we consider $l=7$ for the electric basis and $l=3$, $L=200$ for the magnetic, obtaining a total of five pairs of points.

\subsection{Towards defining a physical scale}
\label{subsec:physscale}
We are not aware of any real experiment which can be described by the effective $(2+1)$-dimensional compact pure gauge theory considered in this paper. Therefore, we cannot extract the physical value of the lattice spacing with large-volume MC calculations\footnote{See Section V C in Ref.~\cite{PhysRevD.106.114511} for the illustration of the principle to determine the value of the lattice spacing.}.
Thus, for the sake of demonstrating our method, we consider an artificial value for the lattice spacing, e.g. $a=0.1~\text{fm}$, and we use the data in the previous sections to identify the physical value for the scales. With two sets of distances, we have two scale factors \textit{s} to connect $r_1$ and $r_2$, ($r_2=s\cdot r_1$), i.e. $r_2=\sqrt{5}\cdot r_1$ and $r_2=\sqrt{\frac{5}{2}}\cdot r_1$. We then combine the results in a single plot.

\begin{figure}[htp!]
    \centering
    \includegraphics[width=1.0\columnwidth]{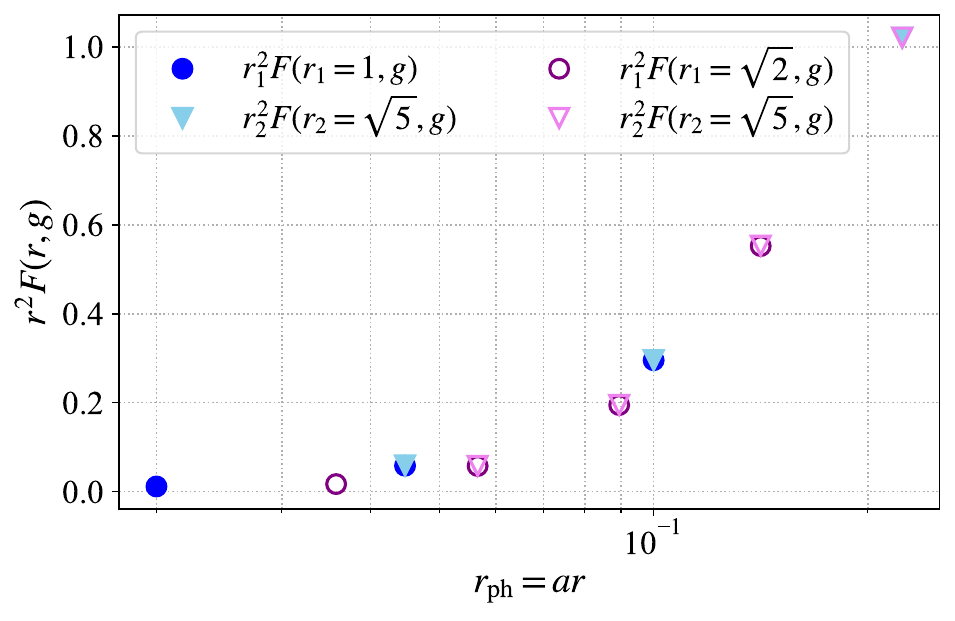}
    \caption{\textbf{Step scaling exact diagonalization results (electric basis and $\bm{l=1}$) as a function of physical distances:} data for set of static forces $F(r_1=1,g)$ ($F(r_2=\sqrt{5},g)$) displayed as \textit{full circles} (\textit{full downward-pointing triangles}) and for $F(r_1=\sqrt{2},g)$ ($F(r_2=\sqrt{5},g)$) displayed as \textit{empty circles} (\textit{empty downward-pointing triangles}).}
    \label{fig:l1physscale_ed}
\end{figure}

Let us first consider the set $r_1=1$, $r_2=\sqrt{5}$, $r_3=\sqrt{8}$. Our aim is to start with $\beta_{MC}$ and invert the sequence by changing the scale by \textit{s} and include the physical value of the lattice spacing, $a=0.1~\text{fm}$. At $\beta_{MC}\equiv \beta_N$ we have,
\begin{subequations}
\begin{align}[left =\beta_N \mapsto \empheqlbrace\,]
    r_{2,\text{ph}}&=r_2\cdot a=0.223~\text{fm},\\
    r_{1,\text{ph}}&=r_1\cdot a=0.1~\text{fm}\label{subeq:r1betan}.
\end{align}
\end{subequations}
Then, we go to the next value of the bare coupling, where we have,
 \begin{subequations}
\begin{align}[left =\beta_{N-1} \mapsto \empheqlbrace\,]
    r_{2,\text{ph}}&=r_2\cdot a/s=  0.1~\text{fm},\label{subeq:r2betanm1}\\
    r_{1,\text{ph}}&=r_1\cdot a/s= 0.045~\text{fm}.
\end{align}
\end{subequations}
The procedure iterates through multiple steps, and eventually, the static force values can be written in terms of a physical scale, as depicted in Fig.~\ref{fig:l1physscale_ed}, with data from Fig.~\ref{fig:stepl1light},\ref{fig:stepl1purple}. Note, for example, that Eqs.~\eqref{subeq:r1betan}, \eqref{subeq:r2betanm1}, correspond to the same physical scale (rightmost \textit{full circle} and second rightmost \textit{full downward-pointing triangle}).

In compact pure gauge $U(1)$ theory, the $\beta$-function is trivial and therefore there is no renormalization of the coupling. Consequently, there is, in principle, no scale dependence.
Nevertheless, in Fig.~\ref{fig:l1physscale_ed}, we observe a non-trivial behavior of the dimensionless quantity $r^2 F(r,g)$ as a function of the physical distance. We attribute this dependence to the following argument:
The bare coupling $g^2$ has a dimension of a mass and we can define a dimensionless coupling $\tilde{g}^2=g^2/\mu$, where $\mu$ has a dimension of a mass. As a consequence, the  $\beta$-function, employing $\tilde{g}^2$, becomes proportional to $-\tilde{g}^2$ see Eq.(1) and Eq.(2) in Ref.~\cite{Raviv:2014xna}.
Therefore the dimensionless quantity $r^2 F(r,g)$, which is proportional to $\tilde{g}^2$, should also show a non-trivial behavior as a function of $\tilde{g}^2$ and, hence, as a function of the physical distance, too.
We successfully tested the $\tilde{g}^2$ behavior of $r^2 F(r,g)$ through a linear fit $c_1+c_2 g^2$ where we absorb the scale $\mu$ in the coefficient $c_2$.
That being the case, $r^2 F(r,g)$ can be used to demonstrate the here proposed step scaling approach.
Note that, when including matter fields, the $\beta$-function becomes non-trivial, see again Ref.~\cite{Raviv:2014xna}. 

\begin{figure}[htp!]
    \centering
    \includegraphics[width=1.0\columnwidth]{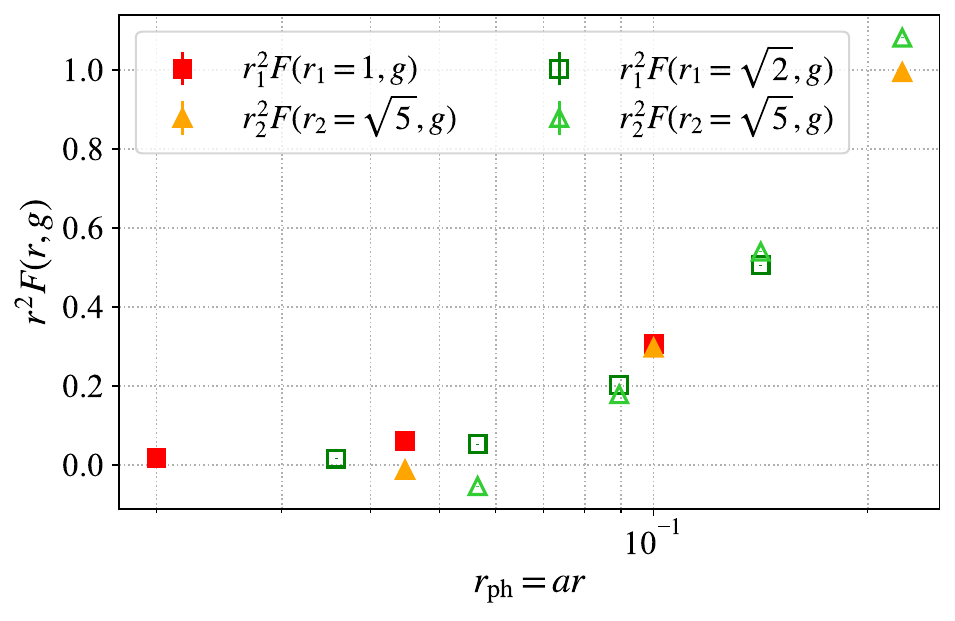}
    \caption{\textbf{Step scaling VQE results (electric basis and $\bm{l=1}$) as a function of physical distances:} data for set of static forces $F(r_1=1,g)$ ($F(r_2=\sqrt{5},g)$) displayed as \textit{full squares} (\textit{full upward-pointing triangles}) and for $F(r_1=\sqrt{2},g)$ ($F(r_2=\sqrt{5},g)$) displayed as \textit{empty squares} (\textit{empty upward-pointing triangles}). The error bars are smaller than the markers.}
    \label{fig:l1physscale_vqe}
\end{figure}

We can replicate the procedure using the outcomes from the variational quantum algorithm (again Figs.~\ref{fig:stepl1light}, \ref{fig:stepl1purple}), as depicted in Fig.~\ref{fig:l1physscale_vqe}. Despite fluctuations in the results, attributed in part to the finite number of shots and the limited convergence of the optimization, the data effectively captures the dependence of the coupling as a function of the physical distance.

\begin{figure}[htp!]
    \centering
    \includegraphics[width=1.0\columnwidth]{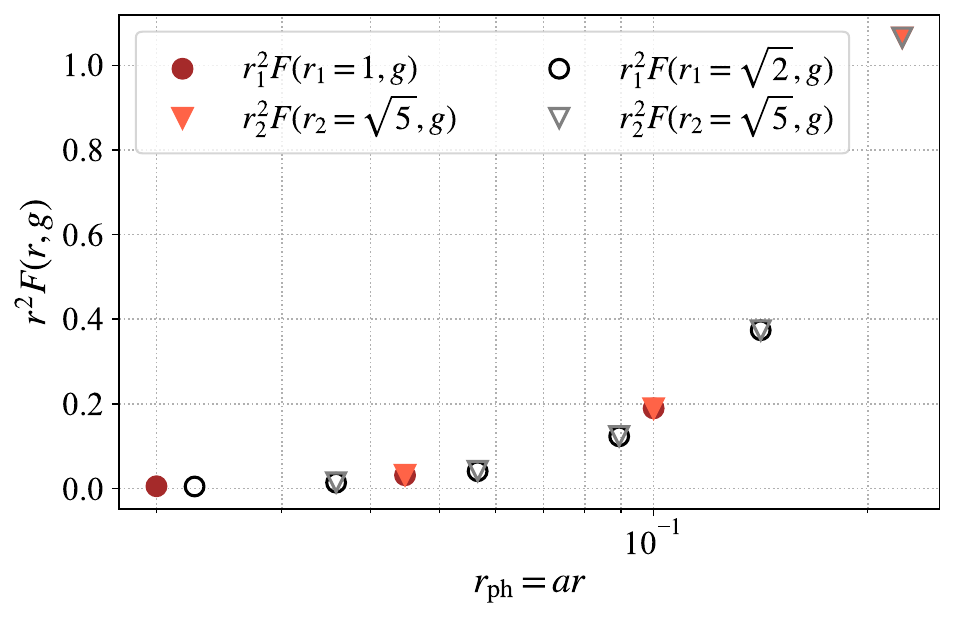}
    \caption{\textbf{Step scaling exact diagonalization results (electric basis $\bm{l=7}$ and magnetic basis $\bm{l=3,L=200}$) as a function of physical distances:} data for set of static forces $F(r_1=1,g)$ ($F(r_2=\sqrt{5},g)$) displayed as \textit{full circles} (\textit{full downward-pointing triangles}) and for $F(r_1=\sqrt{2},g)$ ($F(r_2=\sqrt{5},g)$) displayed as \textit{empty circles} (\textit{empty downward-pointing triangles}).}
    \label{step_l7phys}
\end{figure}

The procedure is repeated also for the analysis with electric and magnetic basis, using the data from Figs.~\ref{step_l7lightblue2}, \ref{step_l7purple2} and combining them in Fig.~\ref{step_l7phys}.

\section{Discussion}\label{sec:discussion}
In a previous work~\cite{PhysRevD.106.114511}, part of the authors outlined the idea of computing the running of the coupling and the $\Lambda$-parameter through a step scaling approach in $(2+1)$-dimensional QED. To this end, the combination of MC and quantum computing methods was proposed. 

Here we performed the very first step towards this final goal by analyzing a compact $U(1)$ gauge theory. 
We designed tailored quantum circuits for the implementation of gauge degrees of freedom. We showed that our program, based on a variational quantum simulation, can be carried out. This is illustrated in Fig.~\ref{fig:l1physscale_vqe}, which shows the dimensionless quantity $r^2F(r,g)$, related to the renormalized coupling, as a function of a physical scale. We remark, that at the moment an artificial value of the lattice spacing was employed. 
Our results provide a successful test for the capability of variational quantum simulations for studying the step scaling function on the lattice, with the potential for further applications of this approach in the future, in particular, with existing and emerging quantum hardware.

The Hamiltonian formalism, discussed in this work, can be related to the action formalism by taking the continuum limit in time of the action with a fixed physical condition.
This allows us to determine the relation of the bare couplings in both cases, by matching, e.g., the corresponding plaquette expectation values. 
In this work, we used a preliminary value of the bare coupling for this matching, obtained by Monte Carlo simulations for a periodic $3 \times 3$ system in Refs.~\cite{Funcke:2022opx,Funcke:2024}.
On the quantum computing side, we were able to achieve results for the expectation value of the plaquette within the same range of bare couplings, see Fig.~\ref{plaqpbcvqe}. In the same figure, it is demonstrated that, employing exact diagonalization, a truncation $l=3$ is sufficient to see convergence. There, we also provided VQE results for a truncation $l=1$, which could be simulated with current quantum resources.

In Appendix~\ref{app:engap}, through a detailed finite-size MC study, we demonstrated that with a system size of $6\times 6$, matching with the mass gap becomes possible, which opens the road for future quantum computations.

In Appendix~\ref{app:matter}, we discuss the theoretical details of the fermionic Hamiltonian, building the ground for future calculations including matter fields. This will then lead to a situation where the running of the coupling is non-trivial and thus can provide a meaningful value of the QED $\Lambda$-parameter in $(2+1)$ dimensions.

In the work presented here, we set the basis of variational quantum simulation of $(2+1)$-dimensional QED. The methodology developed here can be utilized for extensions of the theory, including e.g. topological terms or non-zero matter density and even an analysis of real time evolution.

\begin{acknowledgments}

We thank Gregoris Spanoudes for pointing out the phase factor in the kinetic term in Eq.~\eqref{eq:hkin}. We also thank Angus Kan for the data for the plaquette expectation value at $l=4$ in Fig.~\ref{plaqpbcvqe}.
We are also grateful to Paulo Itaborai for his contribution to the tests of the variational Ansatz.
We thank Christiane Groß for communicating her Monte Carlo results prior to publication.
We want to thank Cristina Diamantini, Fernanda Steffens and Georgios Polykratis for helpful discussions. 

A.C.\ is supported in part by the Helmholtz Association —“Innopool Project Variational Quantum Computer Simulations (VQCS).”
This work is supported with funds from the Ministry of Science, Research and Culture of the State of Brandenburg within the Centre for Quantum Technologies and Applications (CQTA). 
\begin{center}
    \includegraphics[width = 0.08\textwidth]{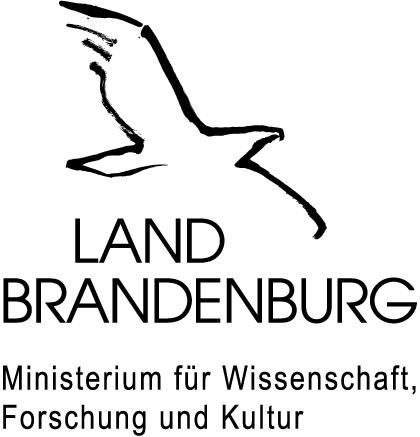}
\end{center}
This work is funded by the European Union’s Horizon Europe Frame-work Programme (HORIZON) under the ERA Chair scheme with grant agreement no. 101087126.
This project was funded by the Deutsche Forschungsgemeinschaft (DFG, German Research Foundation) as part of the CRC 1639 NuMeriQS -- project no.\ 511713970.

PS acknowledges support from: Europea Research Council AdG NOQIA;  MCIN/AEI (PGC2018-0910.13039/501100011033,  CEX2019-000910-S/10.13039/501100011033, Plan National FIDEUA PID2019-106901GB-I00, Plan National STAMEENA PID2022-139099NB, I00,project funded by MCIN/AEI/10.13039/501100011033 and by the “European Union NextGenerationEU/PRTR" (PRTR-C17.I1), FPI); QUANTERA MAQS PCI2019-111828-2);  QUANTERA DYNAMITE PCI2022-132919, QuantERA II Programme co-funded by European Union’s Horizon 2020 program under Grant Agreement No 101017733); Ministry for Digital Transformation and of Civil Service of the Spanish Government through the QUANTUM ENIA project call - Quantum Spain project, and by the European Union through the Recovery, Transformation and Resilience Plan - NextGenerationEU within the framework of the Digital Spain 2026 Agenda; Fundació Cellex; Fundació Mir-Puig; Generalitat de Catalunya (European Social Fund FEDER and CERCA program, AGAUR Grant No. 2021 SGR 01452, QuantumCAT \ U16-011424, co-funded by ERDF Operational Program of Catalonia 2014-2020); Barcelona Supercomputing Center MareNostrum (FI-2023-1-0013); Views and opinions expressed are however those of the author(s) only and do not necessarily reflect those of the European Union, European Commission, European Climate, Infrastructure and Environment Executive Agency (CINEA), or any other granting authority.  Neither the European Union nor any granting authority can be held responsible for them (EU Quantum Flagship PASQuanS2.1, 101113690, EU Horizon 2020 FET-OPEN OPTOlogic, Grant No 899794),  EU Horizon Europe Program (This project has received funding from the European Union’s Horizon Europe research and innovation program under grant agreement No 101080086 NeQSTGrant Agreement 101080086 — NeQST); ICFO Internal “QuantumGaudi” project; European Union’s Horizon 2020 program under the Marie Sklodowska-Curie grant agreement No 847648; “La Caixa” Junior Leaders fellowships, La Caixa” Foundation (ID 100010434): CF/BQ/PR23/11980043.

\end{acknowledgments}

\appendix
\section{Gray encoding variational circuits}
\label{app:gray}
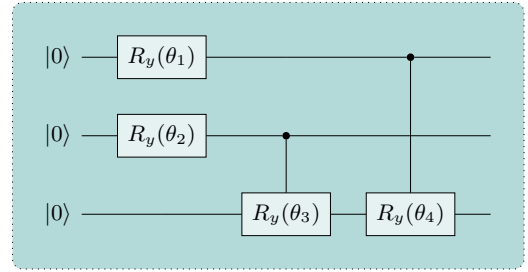
\begin{figure}[H]
\centering
\tikzset{operator/.append style={fill=teal!10}}  
\noindent   \resizebox{0.8\columnwidth}{!}{
 \begin{quantikz}[thin lines] 
 \gategroup[3,steps=6,style={dotted,rounded corners,fill=teal!30,inner sep=10pt},background]{} 
& \lstick{$\ket{0}$} &\gate{R_y(\theta_1)} & \qw & \ctrl{2} & \qw \\
& \lstick{$\ket{0}$} &\gate{R_y(\theta_2)} & \ctrl{1} & \qw & \qw \\
& \lstick{$\ket{0}$} & \qw & \gate{R_y(\theta_3)} & \gate{R_y(\theta_4)} & \qw     
\end{quantikz}
}   
\caption{\textbf{Variational circuit for Gray encoding with $\bm{l=3}$:} $\ket{010}$ represents the \textit{vacuum state} and the state $\ket{100}$ excluded.}
\label{grayl3}
\end{figure}

As outlined in Section~\ref{sec:Numerical_setup}, the qubit requirements for the adopted encoding follow a logarithmic scaling. Specifically, for certain values of $l$, the exclusion of a single state suffices and as a consequence, the complexity of the required gate set for circuit implementation is significantly reduced.
The logic of the exclusion of states outside the reduced Hilbert space $\mathcal{H}_{2l+1}$ for $l>1$ mirrors that of the $l=1$ case. Moreover, when only one state requires exclusion, a discernible pattern emerges in the gate structure.
In this section, we present the variational circuits corresponding to $l=3,7,15$, in Figs.~\ref{grayl3},~\ref{grayl7},~\ref{grayl15} respectively.

\begin{figure}[H]
\centering
\tikzset{operator/.append style={fill=violet!10}}  
\noindent   \resizebox{0.9\columnwidth}{!}{
 \begin{quantikz}[thin lines] 
 \gategroup[4,steps=7,style={dotted,rounded corners,fill=violet!30,inner sep=10pt},background]{} 
& \lstick{$\ket{0}$} &\gate{R_y(\theta_1)} & \qw & \qw & \ctrl{1} & \qw \\
& \lstick{$\ket{0}$} &\gate{R_y(\theta_2)} & \qw & \ctrl{2} & \gate{R_y(\theta_6)} &\qw \\
& \lstick{$\ket{0}$} &\gate{R_y(\theta_3)} & \ctrl{1} & \qw & \qw & \qw \\
& \lstick{$\ket{0}$} & \qw & \gate{R_y(\theta_4)} & \gate{R_y(\theta_5)} & \qw & \qw
\end{quantikz}
}   
\caption{\textbf{Variational circuit for Gray encoding with $\bm{l=7}$:}  $\ket{0100}$ represents the \textit{vacuum state} and the state $\ket{1000}$ excluded.}
\label{grayl7}
\end{figure}
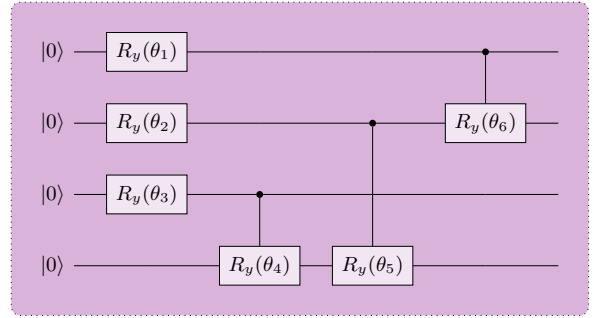
\begin{figure}[H]
\centering
\tikzset{operator/.append style={fill=orange!10}}  
\noindent   \resizebox{0.9\columnwidth}{!}{
 \begin{quantikz}[thin lines] 
 \gategroup[5,steps=8,style={dotted,rounded corners,fill=orange!30,inner sep=10pt},background]{} 
& \lstick{$\ket{0}$} &\gate{R_y(\theta_1)} & \qw & \qw & \qw &\ctrl{1} & \qw \\
& \lstick{$\ket{0}$} &\gate{R_y(\theta_2)} & \qw & \qw & \ctrl{1} & \gate{R_y(\theta_8)}& \qw \\
& \lstick{$\ket{0}$} &\gate{R_y(\theta_3)} & \qw & \ctrl{2} & \gate{R_y(\theta_7)} &\qw &\qw \\
& \lstick{$\ket{0}$} &\gate{R_y(\theta_4)} & \ctrl{1} & \qw & \qw & \qw &\qw \\
& \lstick{$\ket{0}$} & \qw & \gate{R_y(\theta_5)} & \gate{R_y(\theta_6)} & \qw & \qw &\qw
\end{quantikz}
}   
\caption{\textbf{Variational circuit for Gray encoding with $\bm{l=15}$:}  $\ket{01000}$ represents the \textit{vacuum state} and the state $\ket{10000}$ excluded.}
\label{grayl15}
\end{figure}
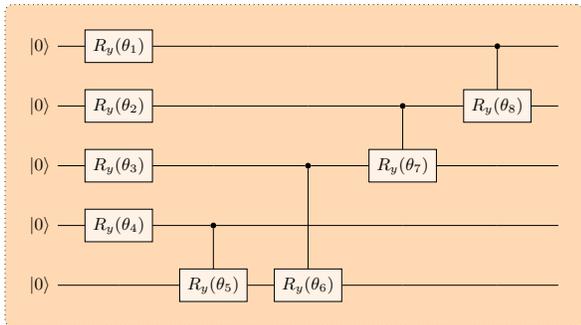
The next step is to establish an appropriate pattern for entangling multiple gauge fields. This task is particularly important when working with an electric (magnetic) basis formulation, especially in regimes characterized by weak (strong) coupling. 
In this paper, we devised a structure capable of entangling all gauge fields, while guaranteeing the exclusion of unphysical states. 

We remark that integrating Gauss's law a priori makes the Hamiltonian non-local, and hence requires more gates per qubit. See e.g. Ref~\cite{carena2024quantum} for a discussion on the effect on a real noisy quantum device. 
In this work, the noise is purely statistical, as we simulate classically the behavior of a quantum device.
In the future, our work will be devoted to exploring more resource-efficient alternatives. 
This will involve an analysis of an efficient mapping of the square lattice structure to quantum devices with limited connectivity.

The Python code produced in this study is available at Ref.~\cite{QEDHamiltrepo}.

\section{\boldmath$\Delta E$}\label{app:engap}
In this section, we report our investigation of the mass gap $\Delta E$ with Monte Carlo simulations.
We find numerical evidence that the volume $L/a=3$ used with the present Hamiltonian formulation is too small to match the mass gap of the theory.
Nonetheless, we provide some quantitative estimate of the coupling range needed for this approach, finding that a slight increase in the volume size on the Hamiltonian simulations would allow for such a matching procedure.

A $U(1)$ gauge theory in $(2+1)$-dimensions is a non-trivial theory only at finite lattice spacing,
where it can be approximated by a plasma of magnetic monopoles in an external field~\cite{POLYAKOV197582}.
It has been shown that the theoretical prediction for the gauge coupling dependence of the mass gap is in agreement also with numerical determinations~\cite{KARLINER1983371}.
In the continuum limit, the theory becomes equivalent to a free massive scalar theory~\cite{Gopfert:1981er},
whose lightest state is the analogue of a ``massive photon''.
At finite lattice spacing, this is found to be a glueball, with $J^{PC}=0^{--}$ quantum numbers~\cite{Loan:2003wy,Loan:2006gq,Athenodorou:2016ebg,Athenodorou:2018sab,Athenodorou:2018sab}.
The wavefunction of the $0^{--}$ state can be interpolated by the operator~\cite{PhysRevD.59.014512}:
\begin{equation}
    \hat{\phi}(t) = 
    \frac{1}{V}
    \sum_{\vec{r}} \operatorname{Im} \operatorname{Tr} 
    \left[ \hat{P}(t, \vec{r}) - \hat{P}^\dagger(t,\vec{r})\right]
    \, .
\end{equation}
The sum over the spatial coordinates $\vec{r}$ ensures the particle is at rest.
In a Monte Carlo simulation with time extent $T$ we can find the mass gap from the expectation value (over the gauge configurations) of the following correlation function~\cite{gattringer2009quantum},
\begin{equation}
    C_T(t) = \langle \hat{\phi}(t) \hat{\phi}(0) \rangle_T \, .
\end{equation}
We note that $\hat{\phi}$ is hermitian by construction.
According to its spectral decomposition, 
for large values of $t$ the above correlator approaches the expression:
\begin{equation}
    C_T(t) \to |\langle 0^{--}|\hat{\phi}(0)|0 \rangle |^2 \left( e^{-M t} + e^{-M(T-t)} \right) 
    \, ,
\end{equation}
where $\ket{0}$ is the interacting vacuum and we have taken into account also the backward signal from $T-t$.
$M$ is the mass of the lightest state interpolated by the operator $\hat{\phi}$.
For each ensemble, i.e. volume and coupling constant, 
we compute the following effective mass curve for the $0^{--}$ correlator from the following implicit expression,
\begin{equation} \label{eq:MeffCurve}
    \frac{C(t)}{C(t+1)} = 
    \frac{\cosh{(M_\text{eff}(t) \cdot (t - T/2))}}{\cosh{(M_\text{eff}(t) \cdot (t + 1 - T/2))}}
    \, .
\end{equation}
Namely, for each $t$ the value $M_{\text{eff}}(t)$ is found by numerically solving Eq.~\eqref{eq:MeffCurve}.
Finally, we consider the time interval where $M_{\text{eff}}(t)$ plateaus within the statistical uncertainty,
and fit it to a constant value.

We have performed Monte Carlo simulations in the range $1.35\leq \beta\leq 2.25$ and $6\leq L/a\leq 16$.
A software implementation for reproducing the Monte Carlo simulations is available at Ref.~\cite{Urbach_Gross_Romiti}.
In our setup $T/a=16$ for all the ensembles. We have checked that this value of the time extent is compatible with the infinite time extent limit within the uncertainty.
Similarly, we have verified that in this range of $\beta$, the values at $L/a=16$ are compatible with their infinite volume limit.

The gauge field configurations have been produced using the Hybrid Monte Carlo (HMC) algorithm (see Ref.~\cite{Duane:1987de}).
The effects of autocorrelation have been taken into account with the method of Ref.~\cite{Wolff:2003sm} in order to correctly estimate the uncertainty on the data.

We recall that ultimately our goal consists of matching the Lagrangian and Hamiltonian results,
with the latter being limited to $L/a=3$.
A direct evaluation of the mass gap from euclidean correlators at this volume is challenging, 
as the excited states contamination becomes significant and the signal-to-noise-ratio of the correlator becomes poor.
Therefore, we are interested in the small volume extrapolations.
\begin{figure}[htp!]
    \centering
    \includegraphics[width=1.0\columnwidth]{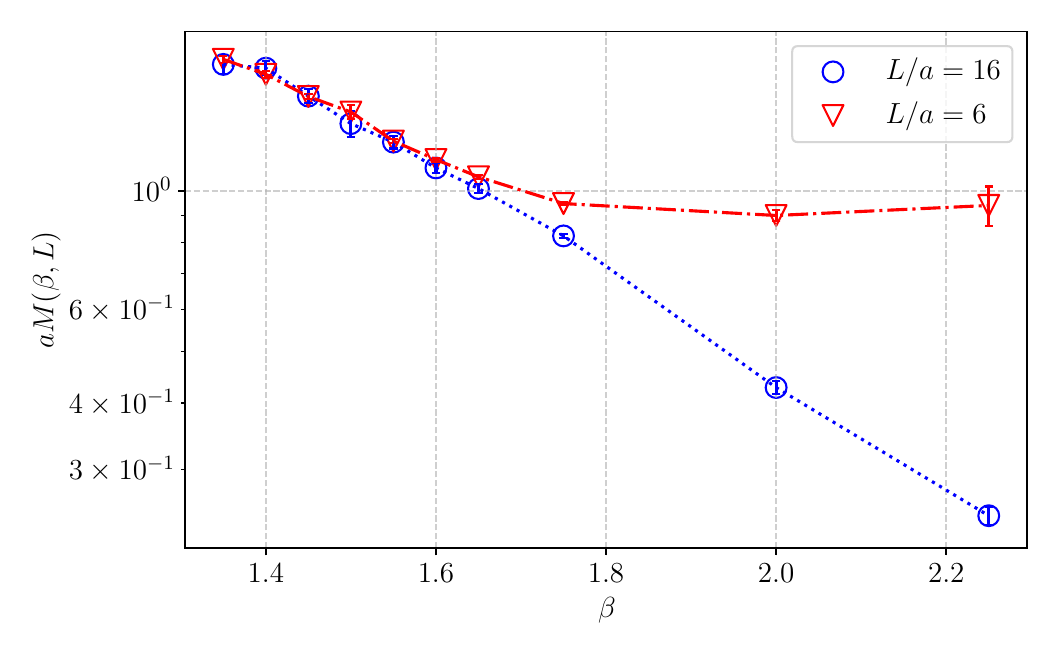}
    \caption{\textbf{$\bm{\beta}$ dependence of the glueball mass with quantum numbers $\bm{J^{PC}=0^{--}}$:}
    Different lines correspond to different values of $L/a$ in the legend and $T/a=16$ for all points.}
    \label{fig:MglueballBeta}
\end{figure}

\begin{figure}[htp!]
    \centering
    \includegraphics[width=1.0\columnwidth]{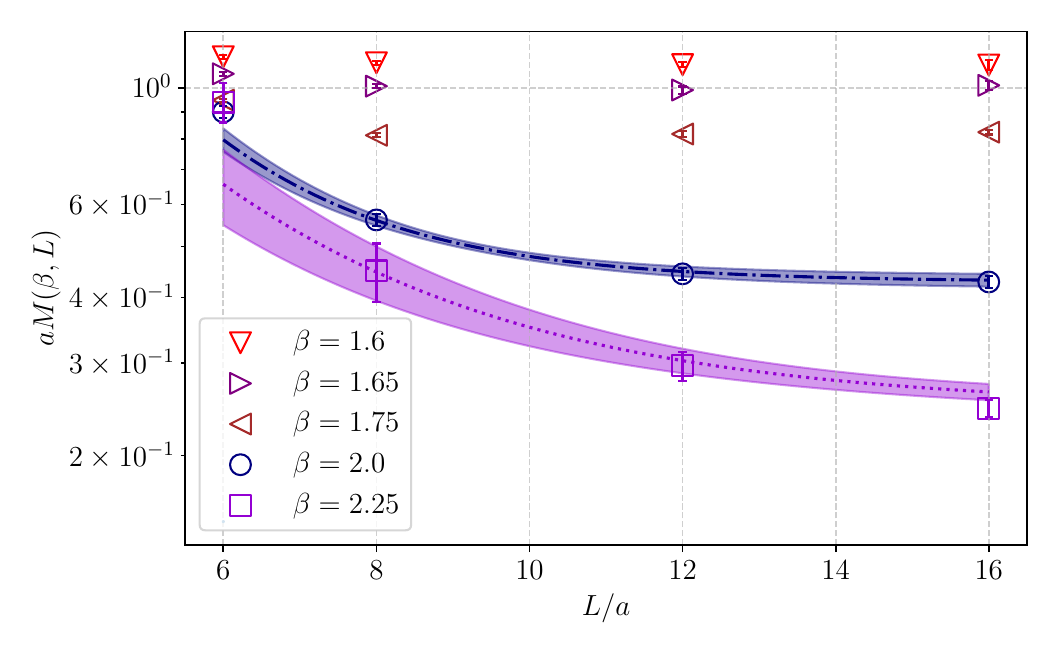}
    \caption{\textbf{Volume dependence of the glueball mass with quantum numbers $\bm{J^{PC}=0^{--}}$:}
    Different lines correspond to different values of $\beta$ in the legend.
    The plot is limited to the values of $\beta \geq 1.6$ for better visualization of FVEs. 
    The fit from Eq.~\eqref{eq:FVEMass} for the two highest values of $\beta$ and $L/a \geq 8$ is shown explicitly.
    }
    \label{fig:MglueballVolume}
\end{figure}

In Fig.~\ref{fig:MglueballBeta}, we show the $\beta$ dependence of the mass of the glueball $0^{--}$,
while in Fig.~\ref{fig:MglueballVolume} we show the volume dependence of the latter.
Finite Volume Effects (FVEs) are fitted according to the following Ansatz inspired by QCD~\cite{DeGrand:1986sg}:
\begin{equation}
\label{eq:FVEMass}
    M(L) = M(L=16a)
    \left[1 + c \cdot \frac{e^{- \frac{\sqrt{3}}{2}  M(L=16a) \cdot L}}{M(L=16a) \cdot L} \right]
    \, ,
\end{equation}
where we consider the values at $L=16a$ as an approximation of the ones at $L=\infty$, and $c$ is a free parameter of the fit.

\begin{figure}[htp!]
    \centering
    \includegraphics[width=1.0\columnwidth]{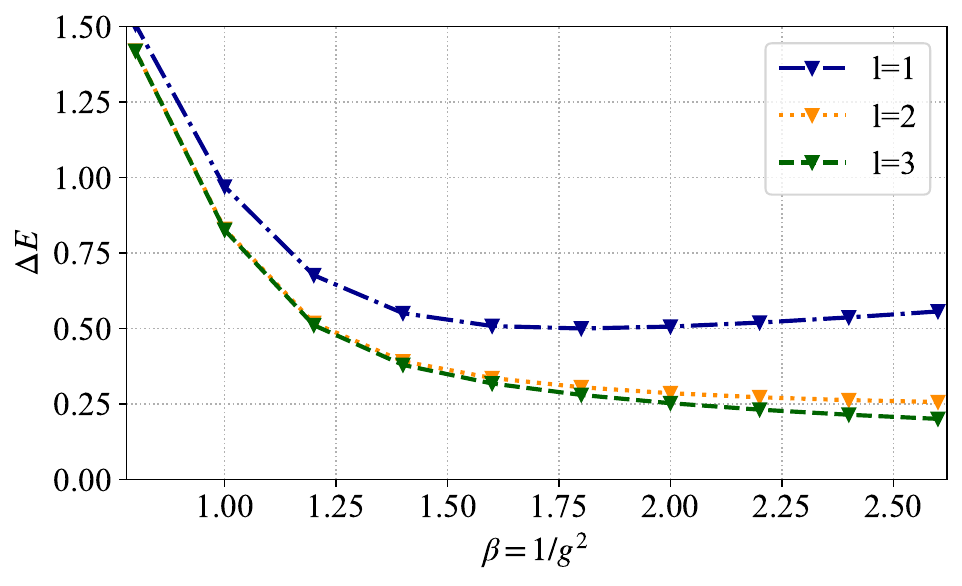}
    \caption{\textbf{Energy gap for $\bm{3\times 3}$ PBC system: }Exact diagonalization results in the Hamiltonian formalism with truncation $l \in [1,3]$.}
    \label{engappbc}
\end{figure}

Looking at the data, we can make the following considerations.
For large $\beta$ we have a smaller lattice spacing, and hence smaller volume in physical units.
As the volume is reduced to $L/a=3$, the Compton wavelength of the glueball does not fit the lattice,
and we cannot interpret the result as the mass of a particle.
Moreover, FVEs are not under control in this region, unless the lattice spacing becomes coarse.
At small $\beta$, the lattice is coarse enough to make the glueball fit, though several values of $\beta$ give values above the cutoff.
According to our results, we conclude that the ``sweet spot'' to perform matching using the glueball mass would be around $\beta\approx 1.8$, and with a Hamiltonian with at least $(L/a)\geq 6$.
It is understood that this value of the coupling should be extrapolated to the Hamiltonian limit~\cite{Funcke:2022opx} in order to compare the two formalisms.
We leave this type of analysis to future work. The steps to match the Hamiltonian to the Lagrangian formulation shall be the same as for the plaquette expectation value, as described in Section~\ref{sec:matching}, but applied to the mass gap. We remark that at the moment we have computed the mass gap in the Hamiltonian formulation with truncations $l \in [1,3]$ by using exact diagonalization, as illustrated in Fig.~\ref{engappbc}. The missing part is the Hamiltonian limit of $\beta_{MC}$ and the glueball mass.

\section{Electric/magnetic basis analysis}
\label{app:elmag}
\begin{figure}[htp!]
\centering
   \includegraphics[width=1.0\columnwidth]{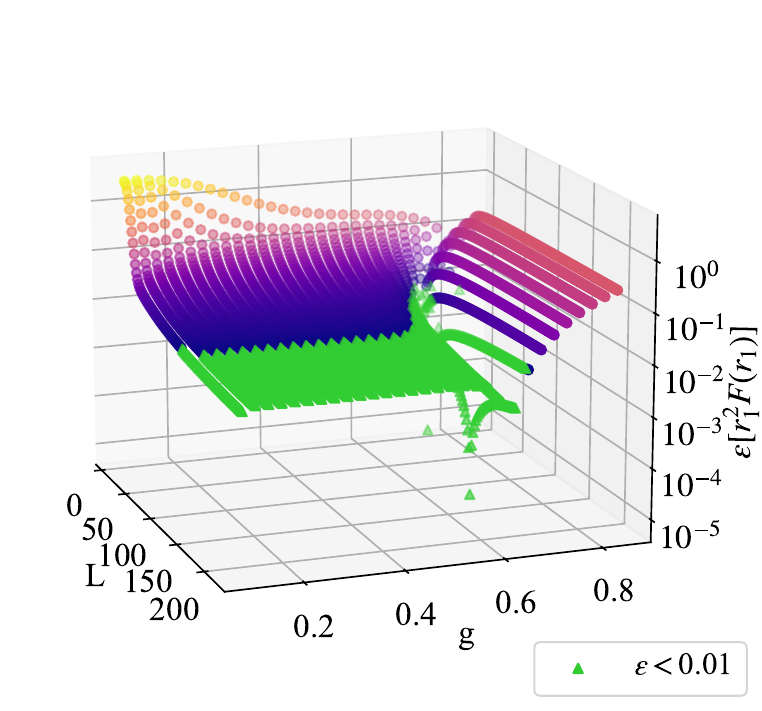}
   \includegraphics[width=1.0\columnwidth]{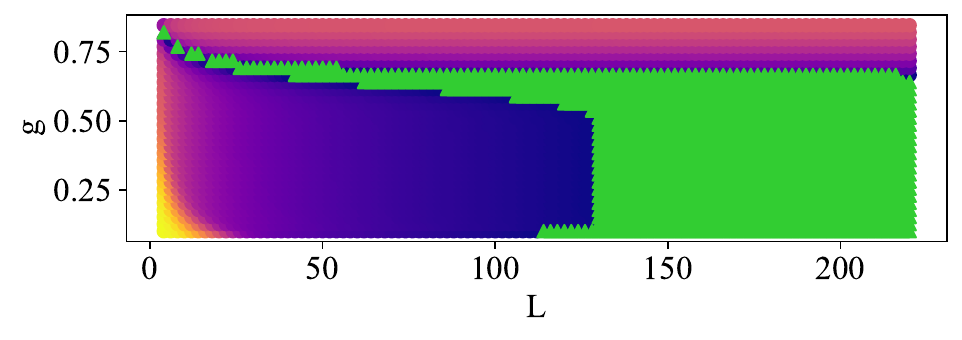}
\caption{(\textit{top}) \textbf{Relative error for static force $\bm{r_1^2F(r_1=1,g)}$ in magnetic basis ($\bm{l=3}$):} the \textit{triangles} correspond to an error below the threshold, $\epsilon<0.01$. (\textit{bottom}) \textbf{Data seen from above:} For small $g$, a larger discretization $L$ is required to reach an accurate result. When increasing $g$, the $L$ parameter must be decreased to capture more states (see text for more details).  }
\label{magl3L200f1}
\end{figure}

\begin{figure}[htp!]
    \centering
    \includegraphics[width=1.0\columnwidth]{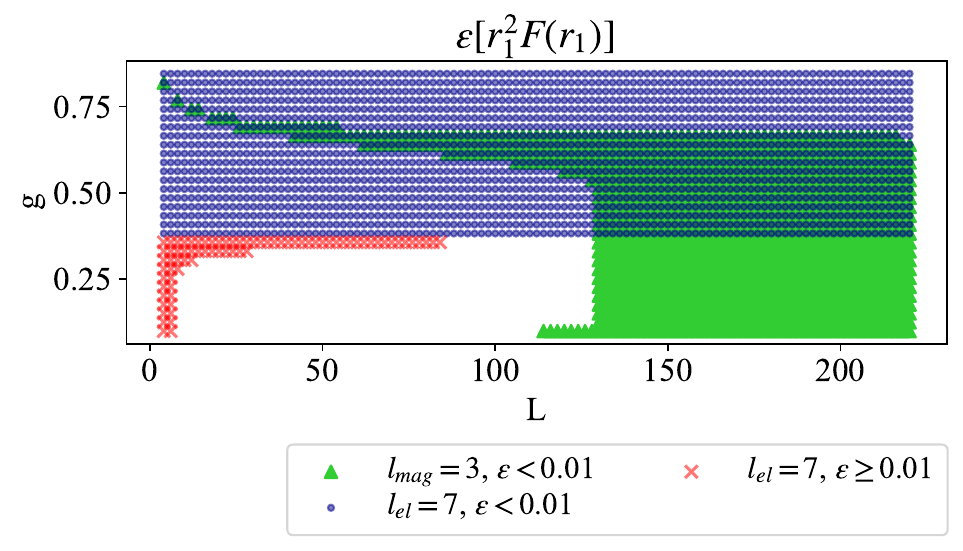}
    \caption{\textbf{Comparison between electric and magnetic basis results for $\bm{r_1^2F(r_1=1,g)}$:} the \textit{small dots} correspond to results with $l_{el}=7$ in the electric basis and with relative errors $\epsilon$ lower than the corresponding ones in the magnetic basis. At $g\sim 0.383$ and below, the electric data have $\epsilon\geq 0.01$ (\textit{crosses}). At this point, the magnetic basis ($l_{mag}=3$) is considered for the computation. For this regime, a discretization parameter $L$ within the \textit{triangles} region is selected.}
    \label{l3lightf1}
\end{figure}

\begin{figure}[htp!]
\centering
   \includegraphics[width=1.0\columnwidth]{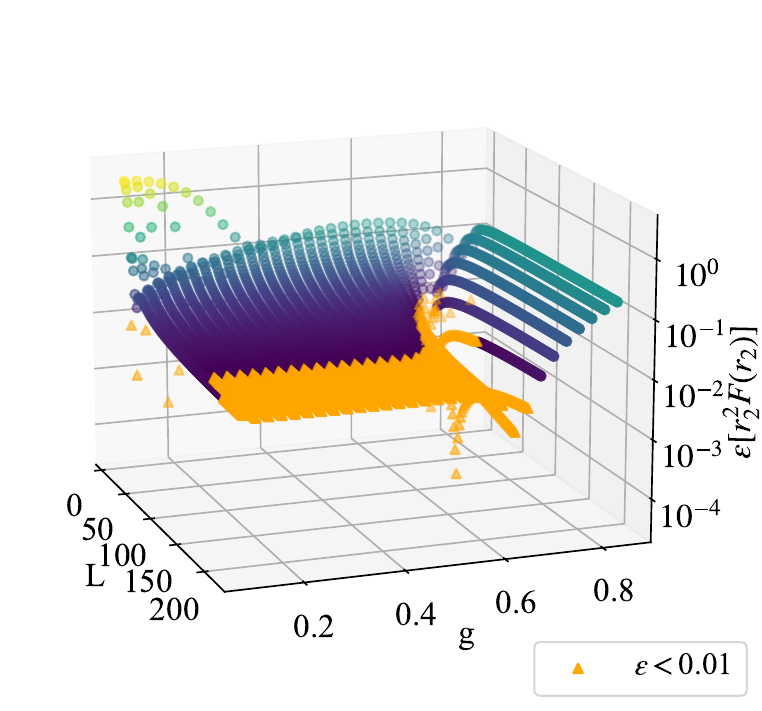}
   \includegraphics[width=1.0\columnwidth]{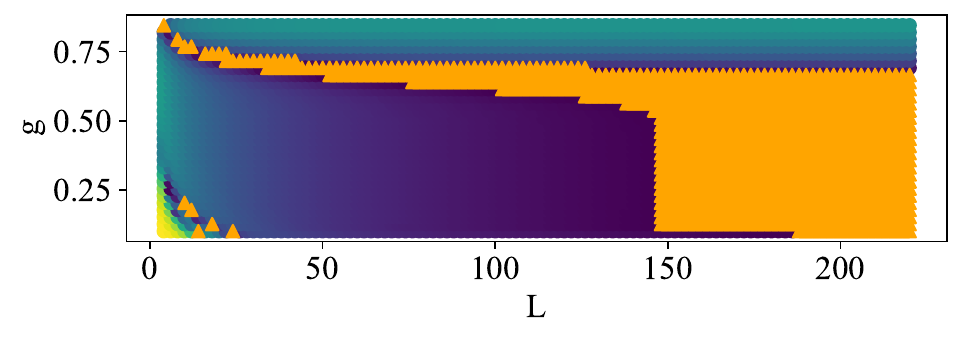}
\caption{(\textit{top}) \textbf{Relative error for static force $\bm{r_2^2F(r_2=\sqrt{5},g)}$ in magnetic basis ($\bm{l=3}$):} the \textit{triangles} correspond to an error below the threshold, $\epsilon<0.01$. (\textit{bottom}) \textbf{Data seen from above:} For small $g$, a larger discretization $L$ is required to reach an accurate result. When increasing $g$, the $L$ parameter must be decreased to capture more states (see text for more details).  }
\label{magl3L200f2}
\end{figure}

\begin{figure}[htp!]
    \centering
    \includegraphics[width=1.0\columnwidth]{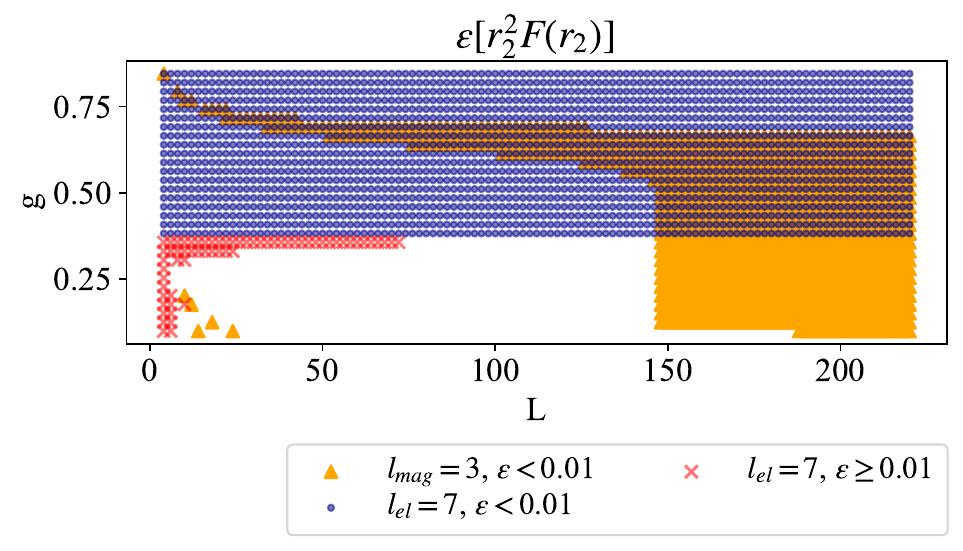}
    \caption{\textbf{Comparison between electric and magnetic basis results for $\bm{r_2^2F(r_2=\sqrt{5},g)}$:} the \textit{small dots} correspond to results with $l_{el}=7$ in the electric basis and with relative errors $\epsilon$ lower than the corresponding ones in the magnetic basis. At $g\sim 0.383$ and below, the electric data have $\epsilon\geq 0.01$ (\textit{crosses}). At this point, the magnetic basis ($l_{mag}=3$) is considered for the computation. For this regime, a discretization parameter $L$ within the \textit{triangles} region is selected.}
    \label{l3lightf2}
\end{figure}

In this appendix, we extend the analysis for the step scaling method of Section~\ref{sec:stepscaling}, with an inclusion of both electric and magnetic basis. We start from the value of $\beta\equiv\beta_{MC}$ where we match with Monte Carlo, and we compute the values of the two static forces both in the electric and magnetic basis, within a range of truncations. For small $\beta$ (large $g$) the electric basis is preferable, as the electric Hamiltonian (see Eq.~\eqref{eq:hel}) is dominant and we do not need a large value of $l$. We decided to consider $l=7$, both to have good precision in the results, but also because with a Gray-type encoding (see Section~\ref{sec:Numerical_setup} in the main text and Appendix~\ref{app:gray}), we can exploit a limited number of qubits for a higher number of states. We also use a larger truncation in the electric basis, i.e. $l=32$, to have a reference value\footnote{The value is chosen by testing the convergence of $l$ up to a precision where the absolute error between the results with $l$ and $l-1$ is $\mathcal{O}(10^{-5})$ or lower.} for comparison with the magnetic basis, in fact with this high truncation the results are more likely independent of $l$ and have a physical meaning. For the magnetic basis, we also fix $l=7$ and scan over the discretization parameter $L$, to find a good approximation of the previous reference value. We then consider a relative error to be $\frac{E_{\text{meas}}-E_{\text{exact}}}{E_{\text{exact}}}\equiv\epsilon<0.01$ (where $E_{\text{meas}}$ is the value of the quantity analyzed with fixed truncation and $E_{\text{exact}}$ is the reference value). In the interval of interest, $1.4\leq\beta\leq 10^2$, we select which, between electric basis ($l=7$) and magnetic basis ($l=7$ and $L$), gives us the lowest relative error. 
The results are then used to determine where to switch between both bases in the step scaling procedure, as described in Section~\ref{sec:stepscaling}. We also tested that even with a smaller truncation $l=3$, and higher discretization, we can reproduce the same results, thus with a lower number of resources, high precision can be still guaranteed. Note that the magnetic basis may have the advantage of needing a lower truncation when $\beta\gg1$, but in general has a more complicated Hamiltonian.
In the following, we will show the data of the static forces $F(r_1=1,g)$ and $F(r_2=\sqrt{5},g)$ with $l=3$, the other case follows a similar procedure. 
The selection method systematically explores the potential values of $L$ with a fixed $l$ for the magnetic basis. In the top panel of Fig.~\ref{magl3L200f1}, we present the relative error $\epsilon$ for the quantity $r_1^2F(r_1=1,g)$ across various ranges of $g$ and $L$ (horizontal axes). The results that satisfy the condition $\epsilon<0.01$ are considered suitable for the scaling approach and are highlighted with \textit{triangles}. Examining the bottom panel, which displays the same data with a color-map, we observe that for small couplings, a discretization of at least $L>130$ is required. Conversely, in the stronger coupling region, the magnetic data demonstrate higher precision only for specific values of $L$. This behavior is in line with expectations, as an increase in $g$ should allow for the inclusion of more states within the truncation, ultimately favouring $L=l+1$. However, the fluctuations are remarkably pronounced and the optimal value of $L$ strongly depends on the considered value of $g$. Given the wide range of couplings scanned during the step scaling process, adjusting the discretization for minor intervals of $g$ is deemed inefficient. Consequently, we opt for the magnetic basis only with a stable configuration of input parameters.
The comparison procedure, depicted in Fig.~\ref{l3lightf1}, is outlined through the following steps: initially, we identify the data in the electric basis ($l_{el}=7$) that yields a relative error at a specific $g$, lower compared to the magnetic basis. Subsequently, we sequentially scan from top to bottom until we determine the minimum bare coupling at which the electric basis remains viable.
This condition is established when the results exhibit a relative error below the fixed threshold of $\epsilon(\text{el.\ basis})<0.01$. At this critical coupling ($g\sim 0.383$), we switch to the magnetic formulation ($l_{mag}=3$) and select a truncation parameter $L$ that satisfies the above condition. Since the dimension of the Hilbert space remains independent of the choice of $L$  we opt for a large discretization value that can accommodate both $r_1^2F(r_1,g)$ and $r_2^2F(r_2,g)$ calculations.
An equivalent analysis can be repeated for the second static force and compute $r_2^2F(r_2=\sqrt{5},g)$, see Fig.~\ref{magl3L200f2} for magnetic basis convergence and Fig.~\ref{l3lightf2} for electric/magnetic comparison.
This analysis was used for the two sets of distances in Section~\ref{subsec:set1} and~\ref{subsec:set2}.


\newpage
\section{Matter fields}
\label{app:matter}
In this appendix, we aim to explain the fermionic part of the QED Hamiltonian.
The starting point is the definition of the Dirac matrices in a $(2+1)$-dimensional theory. In the following, we will consider the metric $g^{\mu \nu} = (+, -)$. We have $\gamma^0$ hermitian and  $\gamma^i$ anti-hermitian satisfying,
\begin{equation}
    (\gamma^0)^2 = \mathbb{1},  \ \  (\gamma^i)^2 = -\mathbb{1}.
\end{equation}
In this work, we have chosen the $\gamma$-matrices as,
\begin{equation}\label{eq:gammas}
\gamma_0 = \sigma_z,
    \ \ \ 
\gamma_1 = -i \sigma_y,
    \ \ \ 
\gamma_2 = -i \sigma_x,
\end{equation}
where $\sigma_i$, $i=x,y,z$ are Pauli matrices.
The fermionic Hamiltonian on a lattice of size $N_x$ ($N_y$) in $x$ ($y$) direction is written as,

\begin{align}
\begin{aligned}
    &H =-i \sum_{\vec{r}=(0,0)}^{(N_x,N_y)} \\ &\Bigg[\bar{\psi}(\vec{r}) \gamma^1 \frac{U_{\vec{r},x}\psi(r_x+1,r_y)-U^\dagger_{\vec{r},-x}\psi(r_x-1,r_y)}{2a}  \\
    &+\bar{\psi}(\vec{r})  \gamma^2 \frac{U_{\vec{r},y}\psi(r_x,r_y+1)-U^\dagger_{\vec{r},-y}\psi(r_x,r_y-1)}{2a} \\
    &+m\bar{\psi}(\vec{r})\psi(\vec{r})\Bigg],
    \end{aligned}
    \label{eq:hpsi}
\end{align}
where the first two terms represent the two covariant derivatives in $x$ and $y$ direction and belong to the kinetic part, the last line defines the mass and $\vec{r}=(r_x,r_y)$ represents the lattice site. Here and in the rest of this section, we omit, for simplicity, the operator notation $\hat{O}\to O$. With a two-components spinor $\psi=(\psi_1,\psi_2)^t$ and a staggered formulation~\cite{PhysRevD.11.395}, Eq.~\eqref{eq:hpsi} can be recast and the fermionic fields are selected to have a single component for each site, with arbitrary choice $\psi_1$ ($\psi_2$) on even (odd) sites or vice-versa.
As a last step, $\psi_1$ and $\psi_2$ can be replaced with a single component field $\phi$, obtaining Eq.~\eqref{eq:hmass} and Eq.~\eqref{eq:hkin}, see main text.

We will now discuss a few properties of this formulation.
Starting from the Dirac equation $(i\slashed{\partial}-m)\psi=0$, we have
\begin{align}
     i\gamma^0 \partial_0 \psi + i\gamma^1 \partial_1 \psi  + i\gamma^2 \partial_2 \psi &=m \psi,
\end{align}
or in the Schr\"odinger form,
\begin{equation}
    i\frac{\partial \psi}{\partial t}=H\psi, \ \ \ H=\gamma^0 \gamma^1 p_1 + \gamma^0 \gamma^2 p_2 + \gamma^0 m,
\end{equation}
where $p_j\to -i\partial_j$.
Explicitly substituting the $\gamma$-matrices from Eq.~\eqref{eq:gammas}, one gets,
\begin{align}
    i\dot{\psi}= -i
\begin{pmatrix}
     0& -1  \\
    -1 & 0
\end{pmatrix} \partial_x \psi-i \begin{pmatrix}
    0 & -i  \\
    i & 0
\end{pmatrix} \partial_y \psi+ m\begin{pmatrix}
     1 & 0  \\
    0 &-1 
\end{pmatrix} \psi,
\end{align}
and in terms of the two components,
\begin{align}\label{dirac3d}
    i\begin{pmatrix}
\dot{\psi_1}(n) \\ \dot{\psi_2} (n) 
\end{pmatrix} = i\partial_x
\begin{pmatrix}
\psi_2 (n) \\ \psi_1(n) 
\end{pmatrix}&+
\partial_y \begin{pmatrix}
-\psi_2 (n) \\ \psi_1(n) 
\end{pmatrix}\nonumber \\
&+m \begin{pmatrix}
\psi_1 (n) \\ -\psi_2(n) 
\end{pmatrix}.
\end{align}
Following Refs.~\cite{PhysRevD.11.395,Kogut1978}, we select only the kinetic part of Eq.~\eqref{dirac3d} and rewrite it as
\begin{align}\label{dirac3d_2}
    \begin{pmatrix}
\dot{\psi_1}(n) \\ \dot{\psi_2} (n) 
\end{pmatrix} = \partial_x
\begin{pmatrix}
\psi_2 (n) \\ \psi_1(n) 
\end{pmatrix}+
\partial_y \begin{pmatrix}
i\psi_2 (n) \\ -i\psi_1(n) 
\end{pmatrix},
\end{align}
or in compact form,
\begin{equation}
    \dot{\psi}=\sigma_x \partial_x \psi -\sigma_y \partial_y \psi.
\end{equation}

\begin{figure}[htp!]
    \centering
    \includegraphics[width=0.15\textwidth]{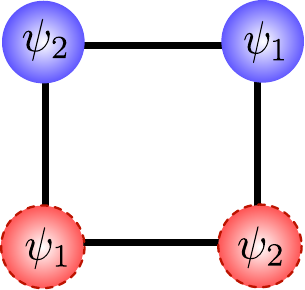}
    \caption{\textbf{Geometric interpretation of Dirac equation for a two-component spinor on a 2D lattice:} Each component appears twice on the unit square, once on a $y = \text{const.} $ plane (\textit{lower sites}) and again on a $y = \text{const.} +a$ plane (\textit{upper sites}).}
    \label{2components}
\end{figure}

We consider the upper components of Eq.~\eqref{dirac3d_2} and place $\psi_1$ on a site. Then shifting one site in the $\pm x$ direction couples $\psi_1$ to $\psi_2$. Similarly, a shift of one site in the $\pm y$ direction results in a coupling between $\psi_1$ to $\psi_2$. Therefore, to yield a coherent geometric interpretation of Eq.~\eqref{dirac3d_2}, the components should be arranged as illustrated in Fig.~\ref{2components}.

Note that each component resides twice within the unit square: once on a $y = \text{const.} $ plane and again on the subsequent $y = \text{const.} +a$ plane. Consequently, in the continuum limit, two fermion fields will emerge within this framework. To distinguish between them, we label the fields on the lower $y = \text{const.} $ plane in Fig.~\ref{2components} as $f_i$, $i=1,2$, and those on the upper $y = \text{const.} +a$ plane as $g_i$, $i=1,2$. In this terminology, the Dirac equation, Eq.~\eqref{dirac3d_2}, transforms into:
\begin{subequations}
\begin{align}[left =\empheqlbrace\,]
    \dot{f}&=\sigma_x \partial_x f - \sigma_y \partial_y g\\
    \dot{g}&=\sigma_x \partial_x g - \sigma_y \partial_y f.
\end{align}
\end{subequations}
Now we distinguish two distinct species: the equation of motion for the sum $u =f + g$ satisfies
\begin{equation}
    \dot{u}=(\sigma_x \partial_x  - \sigma_y \partial_y)u,
\end{equation}
and produces one fermion in the continuum limit. The difference $\tilde{d} = f - g$ satisfies
\begin{align}
    \dot{\tilde{d}}=(\sigma_x \partial_x  + \sigma_y \partial_y)\tilde{d},
\end{align}
which is not a Dirac equation because of the different sign in the last term. However, this can be changed via a unitary transformation,
$d=\sigma_x \tilde{d}$, which gives 
\begin{equation}
    \dot{d}=(\sigma_x \partial_x  - \sigma_y \partial_y)d.
\end{equation}
In summary, this fermion method generates two massless fermion fields in the continuum limit. We are, therefore, free to interpret \textit{u} and \textit{d} as the members of an isodoublet, $(u, d)^t$. 

Another interesting property of a formulation with two-component spinors involves the mass term of the Hamiltonian. In Ref.~\cite{PhysRevD.33.3774} it is described how considering spinors with two components yields parity breaking in the mass term, where a parity transformation is defined as,
\begin{align}
    \begin{pmatrix}
\psi_1(x,y)\\ \psi_2 (x,y)
\end{pmatrix} \xrightarrow{\text{P}}
\begin{pmatrix}
\psi_2 (-x,y)\\ \psi_1 (-x,y)
\end{pmatrix},
\end{align}
and it also acts on the vector fields,
\begin{align}
\begin{pmatrix}
    A_1 (x,y)\\ A_2 (x,y)\end{pmatrix} \xrightarrow{\text{P}}
\begin{pmatrix}
-A_1 (-x,y) \\ A_2 (-x,y)
    \end{pmatrix},
\end{align}
i.e. the $U$ operators will be
\begin{subequations}
    \begin{align}
        U_{n,n+e_x} (x,y)\xrightarrow{\text{P}} U_{n,n+e_x} ^\dagger (-x,y)\\
        U_{n,n+e_y} (x,y)\xrightarrow{\text{P}} U_{n,n+e_y} (-x,y).
    \end{align}
\end{subequations}
As also described in Ref.~\cite{PhysRevD.33.3774}, for a parity-conserving theory, one should then consider a study of four-component spinors. Since this is beyond the purpose of this work, we will not further discuss it.

\bibliography{references}

\end{document}